\documentstyle[aps,preprint]{revtex}
\input epsf
\begin{document}

\title{Charge and Spin Structures of a $d_{x^2 - y^2}$ Superconductor
in the Proximity of an Antiferromagnetic Mott Insulator.}

\author{ F.F. Assaad$^{1,\star}$, M. Imada$^{2}$  and D.J. Scalapino$^{1}$ \\
	 $^{1}$ Department of Physics, University of California,
           Santa Barbara, CA 93106-9530 \\
         $^{2}$ Institute for Solid State Physics, University of Tokyo, 
         7-22-1 Roppongi,
         Minato-ku, Tokyo 106, Japan.  }

\maketitle
\begin{abstract}
To the Hubbard model on a square lattice we add an interaction, $W$, 
which depends upon the square of a near-neighbor hopping. We 
use zero temperature quantum Monte Carlo simulations on lattice 
sizes up to $16 \times 16$, to show that at half-filling and  constant 
value of the Hubbard repulsion, the interaction $W$ triggers a quantum
transition between an antiferromagnetic Mott insulator and  a $d_{x^2 -y^2}$ 
superconductor.  With a combination of finite temperature
quantum Monte Carlo simulations and the Maximum Entropy method, we study
spin and charge degrees of freedom in the superconducting state. 
We give numerical evidence for the 
occurrence  of a finite temperature Kosterlitz-Thouless transition to 
the $d_{x^2 -y^2}$ superconducting state. Above and below the 
Kosterlitz-Thouless transition temperature, $T_{KT}$, we compute 
the 
one-electron density of states, $N(\omega)$, the spin relaxation rate
$1/T_1$, as well as the imaginary and real part of the spin susceptibility
$\chi(\vec{q},\omega)$. The spin dynamics are 
characterized by the vanishing of $1/T_1$ and
divergence of ${\rm Re } \chi( \vec{q} = (\pi,\pi), \omega = 0 )$ in the
low temperature limit. As  $T_{KT}$ is approached
$N(\omega)$ develops a pseudo-gap feature and below $T_{KT}$
${\rm Im } \chi( \vec{q} = (\pi,\pi), \omega )$ shows a peak at finite
frequency.    
\end{abstract}

\section{Introduction}
The motivation of this work is to examine the competition and relationship 
between an antiferromagnetic Mott insulating state  and  a
$d_{x^2 - y^2} $ superconducting state in two dimensions 
using numerical simulations. 
In particular, we are interested in the interplay of these two states 
and the remnant  of the antiferromagnetic correlations in the 
superconducting state. We will see that significant antiferromagnetic 
fluctuations remain in the $d_{x^2 - y^2}$ superconducting state. 
The antiferromagnetic Mott insulator is well described by the
ground state of the half-filled Hubbard model on a square lattice
with on-site Coulomb repulsion $U$  and nearest neighbor single-particle 
hopping $t$ \cite{Scalapino94}.  To this model, we add 
an extra term, $W$, which depends upon the  square
of the single-particle nearest-neighbor hopping.
Staying at half-band filling and constant value of $U$, we have previously 
shown that we can generate a 
quantum transition as a function of the coupling strength, $W$, 
between an antiferromagnetic Mott insulating state and a $d_{x^2 - y^2}$
superconducting state \cite{Assaad97}.
Technically, with the system at half-filling, and with our particular choice of 
$W$, we encounter no fermion sign problem in the Quantum Monte Carlo (QMC)
simulations. Large lattice sizes and low temperatures may thus be reached 
without loss of precision. 
While there are various ways to justify the form of the 
microscopic Hamiltonian, we view the choice of the interaction more 
formally as a means of inducing the  desired quantum transition. In fact,
one of the reasons for choosing this form is that it has a
simple Hubbard-Stratonovitch representation which is  useful in
constructing the Monte Carlo simulation.  Although the formalism with 
the $W$ term may be extended to any lattice structure, we study here the 
two-dimensional system on a square lattice. 

The basic half-filled Hubbard model that we will study has the Hamiltonian
\begin{equation}
      H_U  =  -\frac{t}{2} \sum_{\vec{i}} K_{\vec{i}} + U \sum_{\vec{i}}
         (n_{\vec{i},\uparrow}-\frac{1}{2})
         (n_{\vec{i},\downarrow} -\frac{1}{2}) 
\end{equation}
with the hopping kinetic energy
\begin{equation}
 	K_{\vec{i}} = \sum_{\sigma, \vec{\delta}}
   \left(c_{\vec{i},\sigma}^{\dagger} c_{\vec{i} + \vec{\delta},\sigma} +
        c_{\vec{i} + \vec{\delta},\sigma}^{\dagger} c_{\vec{i},\sigma} \right). 
\end{equation}
Here, 
$c_{\vec{i},\sigma}^{\dagger}$ ($c_{\vec{i},\sigma}$) creates (annihilates) an
electron with {\it z}-component of spin $\sigma$ on site
$\vec{i}$, $n_{\vec{i},\sigma } =  c_{\vec{i},\sigma}^{\dagger}
c_{\vec{i}\sigma}$, and $\vec{\delta} = \pm \vec{a}_x, \pm \vec{a}_y $ where
$\vec{a}_x$, $\vec{a}_y$  are the lattice constants. The energy 
will be measured in units of $t$.  We consider the boundary conditions
\begin{equation}
\label{Bound}
 c_{\vec{i} + L \vec{a}_x, \sigma } = \exp \left(- 2 \pi i \Phi/\Phi_0
\right) c_{\vec{i}, \sigma} $ and $c_{\vec{i} + L \vec{a}_y, \sigma }
= c_{\vec{i}, \sigma},
\end{equation}
with $\Phi_0 = h c / e$ the flux quanta and  $L$ the  linear length
of the square lattice.
The boundary conditions given by Eq. (\ref{Bound})
account for a magnetic flux threading a torus on which  the lattice is
wrapped.

The interaction that we will add has the form:
\begin{equation}
H_W  =  -W \sum_{\vec{i}} K_{\vec{i}}^{2}
\end{equation}
with positive $W$. The Hamiltonian 
\begin{equation}
\label{tUW}
        H = H_U + H_W
\end{equation}
has the possibility of exhibiting a
quantum transition between an antiferromagnetic Mott insulating state and a
superconducting $d_{x^2 - y^2}$  phase. 
When $ W = 0$, the half-filled Hubbard model with
a finite $U$ is known to be a Mott insulator with long-range  antiferromagnetic
order. 
The interaction $H_W$ can be decomposed into  the following terms:
\begin{eqnarray}
\label{HW1}
	H_W = & & H^{(1)}_W + H^{(2)}_W  + H^{(3)}_W + H^{(4)}_W  \nonumber \\
& &H^{(1)}_W  = - 4 W \sum_{\vec{i}, \sigma} c_{\vec{i},\sigma}^{\dagger}
c_{\vec{i},\sigma} \; \; -  \; \;  
W \sum_{\vec{i}, \sigma, \vec{\delta}, \vec{\delta'}} 
c_{\vec{i} + \vec{\delta} ,\sigma}^{\dagger} c_{\vec{i} + \vec{\delta'},\sigma} 
\nonumber \\ & &
H^{(2)}_W =  -  W \sum_{\vec{i}, \sigma, \vec{\delta}, \vec{\delta'}} 
\left( c_{\vec{i},\sigma}^{\dagger} c_{\vec{i},-\sigma}^{\dagger}
       c_{\vec{i}+ \vec{\delta'},-\sigma}c_{\vec{i}+ \vec{\delta},\sigma}  
 + {\rm h.c.} \right)
\nonumber \\ & &
H^{(3)}_W  = +W \sum_{\vec{i}, \vec{\delta}, \vec{\delta'}}
\left(  T^{\dagger}_{\vec{i}, \delta', 1} T_{\vec{i}, \delta , 1} + 
        T^{\dagger}_{\vec{i}, \delta',-1} T_{\vec{i}, \delta ,-1} +
        T^{\dagger}_{\vec{i}, \delta', 0} T_{\vec{i}, \delta , 0} \right)
\nonumber \\ & &
H^{(4)}_W = -W \sum_{\vec{i}, \vec{\delta}, \vec{\delta'}}
\Delta^{\dagger}_{\vec{i}, \delta'} \Delta_{\vec{i}, \delta}.
\end{eqnarray}
Here, 
$ T^{\dagger}_{\vec{i}, \delta, 1} = c_{\vec{i},\uparrow}^{\dagger} 
c_{\vec{i} + \vec{\delta},\uparrow}^{\dagger}$ ,
$ T^{\dagger}_{\vec{i}, \delta,-1} = c_{\vec{i},\downarrow}^{\dagger} 
c_{\vec{i} + \vec{\delta},\downarrow}^{\dagger}$ ,
$ T^{\dagger}_{\vec{i}, \delta, 0} = \left( 
c_{\vec{i},\uparrow}^{\dagger} c_{\vec{i}+\vec{\delta},\downarrow}^{\dagger} +
c_{\vec{i},\downarrow}^{\dagger} c_{\vec{i}+\vec{\delta},\uparrow}^{\dagger} 
\right)/\sqrt{2} $, and
$ \Delta^{\dagger}_{\vec{i}, \delta} = \left(
c_{\vec{i},\uparrow}^{\dagger} c_{\vec{i}+\vec{\delta},\downarrow}^{\dagger} -
c_{\vec{i},\downarrow}^{\dagger} c_{\vec{i}+\vec{\delta},\uparrow}^{\dagger}
\right)/\sqrt{2} $.
$ H^{(1)}_W $ contains single-particle terms which renormalize the 
chemical potential and allow single-particle hopping between next nearest
neighbor sites. $ H^{(2)}_W $ scatters an on-site singlet to adjacent sites.
In the presence of a Hubbard interaction, this term should not contribute
to the low energy physics since double occupancy is suppressed. 
$ H^{(3)}_W $ corresponds to a triplet scattering channel.
Since this term has a positive coupling constant,
and triplet pairing is not favored by the Hubbard interaction, $ H^{(3)}_W $
is not expected to be relevant for the low energy physics.  
Finally $ H^{(4)}_W$  contains the term we are interested in. The terms in
$ H^{(4)}_W$ with $ \delta = \delta' $ contribute to the exchange 
interaction giving
\begin{equation}
W  \sum_{\vec{i}, \vec{\delta} }
 \left(  \vec{S}_{\vec{i}} \cdot \vec{S}_{\vec{i} + \vec{\delta} } -
\frac{1}{4} n_{\vec{i}} n_{\vec{i} + \vec{\delta} }   \right)
\end{equation}
while the terms in $ H^{(4)}_W$  with $  \delta \neq  \delta' $ contribute to a
pairing interaction. Thus in the presence of the on-site Hubbard repulsion,  
$ H^{(4)}_W$  is the relevant part of $H_W$ in determining the low energy 
properties.

As previously  discussed, we have chosen the form of the interaction $H_W$ in 
order of obtaining a model which exhibits a transition from the antiferromagnetic
Mott insulating state to a $d_{x^2 - y^2}$ superconducting state and is suitable 
for Monte-Carlo  simulations. In particular  integrating out the phonons
from a Su-Schrieffer-Heeger \cite{Su} term with Einstein oscillators:
\begin{equation}
	\sum_{\langle\vec{i},\vec{j} \rangle,\sigma} 
    \vec{\lambda} \cdot \left( \vec{Q}_{\vec{i}} - \vec{Q}_{\vec{j}} \right)
   \left(c_{\vec{i}, \sigma}^{\dagger} c_{\vec{j}, \sigma} + {\rm h.c.} \right)
  + \sum_{\vec{i}} \left( \frac{\vec{P}_{\vec{i}}^2}{2M} +
        \vec{Q_{i}}^{\dagger} \frac{D}{2} \vec{Q_{i}} \right),
\end{equation}
and taking the antiadiabatic limit ($M \rightarrow 0 $), generates $H_W$ with 
$ W = \vec{\lambda}^\dagger D^{-1} \vec{\lambda} / 2 $. Pairing mechanism
along those lines were considered in \cite{Hirsch1,Imada1}. Here, however,
we view this simply as a Hubbard Stratonovitch transformation which is 
useful in constructing the Monte-Carlo simulation.

Our general view is that the half-filled two-dimensional Hubbard model is near
various instabilities. In particular, Monte Carlo
calculations find a divergence in the compressibility and an unusually large
dynamical exponent at the metal insulator transition driven by the 
chemical potential.
\cite{Imada,Furukawa,Assaad96}. 
In the doped state obtained with the addition  of a 
chemical potential, the leading
singlet pairing interaction is found in the $d_{x^2 - y^2}$ channel
\cite{Scalapino94,Bulut93}. There is clear
evidence that the model is sensitive to the addition of a chemical potential.
Here, we will examine the half-filled system to the interaction $W$.

The organization of the text is the following. In the next section, we give
a description of the  generalizations required to implement $H_W$ in
the Projector Quantum Monte Carlo (PQMC) 
\cite{Blankenbecler,Koonin,Sandro,Imada89}
and finite 
temperature QMC algorithms \cite{Hirsch85,White}.  In the appendix, we 
discuss  the construction of  trial wave functions used in the PQMC. 
Numerical results are presented in Secs. (\ref{charge}) and
(\ref{spin}). In Sec. (\ref{charge}) we concentrate on the  charge degrees 
of freedom and in Sec. (\ref{spin}) on the spin degrees of freedom. 
The zero temperature data, from which we conclude the existence
of a phase transition between an antiferromagnetic Mott insulating state 
and $d_{x^2-y^2}$ superconducting state has
already been presented in reference \cite{Assaad97}.
At finite temperatures, we concentrate mostly on the superconducting state.
We show the occurrence
of a finite temperature Kosterlitz-Thouless transition. The 
one electron density of states, $N(\omega)$, the relaxation rate 
$1/T_1$,  as well as  the
real and imaginary part of the spin susceptibility  are  calculated  in the 
superconducting phase above and below the Kosterlitz-Thouless transition
temperature.
In the last section, we discuss and summarize our results. 

\section{Quantum Monte Carlo Algorithms}
\label{QMC}

We have applied both the PQMC and  finite temperature QMC algorithms  for
numerical simulations of the Hamiltonian (\ref{tUW}).
The details of those algorithms in the case of the Hubbard model have been
reviewed in the literature \cite{Loh}. In this section, we  discuss the 
generalizations required for the  inclusion of the  $H_W$ term. 
We will concentrate on the description of the  PQMC algorithm. Apart 
from differences in the numerical stabilization of the algorithms, the step 
from the PQMC to finite temperature algorithm is straightforward and
similar to the case of the Hubbard model. For the numerical simulations, 
it is convenient to carry out a canonical transformation, 
$ \tilde{c}_{\vec{i}, \sigma } = 
\exp \left( 2 \pi i \frac{\Phi}{\Phi_0 L}  \vec{i} \cdot \vec{a}_{x} \right)
c_{\vec{i}, \sigma} $. From equation (\ref{Bound}) it follows that 
the  $ \tilde{c} $ fermionic operators satisfy periodic  boundary
conditions. Under this canonical transformation,
\begin{equation}
 	K_{\vec{i}} = \sum_{\sigma, \vec{\delta}}
   \left( \tilde{c}_{\vec{i},\sigma}^{\dagger} 
          \tilde{c}_{\vec{i} + \vec{\delta},\sigma}
           e^{- i \phi \vec{\delta} \cdot \vec{a}_x } + 
        \tilde{c}_{\vec{i} + \vec{\delta},\sigma}^{\dagger} 
        \tilde{c}_{\vec{i},\sigma} 
           e^{  i \phi \vec{\delta} \cdot \vec{a}_x }\right) 
\end{equation}
where $\phi = 2\pi \frac{\Phi}{\Phi_0 L } $.  In this section we will work
in this basis, and omit the {\it tilde} on the fermionic operators
$ \tilde{c} $. The inclusion of magnetic fields in QMC methods is 
straightforward, and has been discussed in \cite{Assaad94b}.

The idea behind the PQMC algorithm is to filter
out the ground state from a trial wave function $| \Psi_T\rangle $ which is
required to be non-orthogonal to the ground state $ | \Psi_0 \rangle $
\begin{equation}
\frac{\langle \Psi_0 | O |  \Psi_0 \rangle }
           {\langle \Psi_0 |  \Psi_0 \rangle }
           = \lim_{ \Theta \rightarrow \infty }
  \frac{ \langle \Psi_T |e^{-\Theta H }
          O
         e^{-\Theta H } | \Psi_T \rangle }
       { \langle \Psi_T |e^{-2\Theta H } | \Psi_T \rangle }
\end{equation}
The Monte Carlo evaluation of the observable $ O$ proceeds in the following
way.
The first step is to carry out a Trotter 
decomposition of the imaginary time propagation:
\begin{equation}
\label{Trotter}
      e^{-2\Theta H  }  \sim
 \left( e^{- \Delta \tau H_U} 
 e^{- \Delta \tau H_W^{(1)} }  \dots 
 e^{- \Delta \tau H_W^{(n_w)} }
        e^{- \Delta \tau H_t } \right)^{m} 
\end{equation}
Here, $H_t$ ($H_U$) denotes the kinetic (potential) 
term of the Hubbard model,  $m \Delta \tau = 2\Theta$ and 
\begin{equation}
H_W = \sum_{n=1}^{n_w} H_W^{(n)}, \; \;
H_W^{(n)} = -W \sum_{r = 1}^{N/n_w}  K_{\vec{k} (n,r) } ^2 
\;  {\rm and}  \;
[K_{\vec{k} (n,r) }, K_{\vec{k} (n,r') } ] = 0 \; \forall \; r, r'. 
\end{equation}
The above defines the function $ \vec{k} (n,r) $,   $ n = 1 \dots n_w$
and $ r = 1 \dots N/n_w $ for an $N$-site  lattice.
Hence, $H_W^{(n)}$ is given by a sum of commuting operators.
The systematic error produced by the above Trotter decomposition is of the 
order  $\Delta \tau$. However, provided that the operators 
$H_W^{(n)}$, $H_t$, $H_U$, $O$, 
as well as the trial wave function $| \Psi_T \rangle $ are simultaneously real
representable the  prefactor of the systematic error proportional to
$\Delta \tau$ in the evaluation of  the observable $O$ vanishes 
\cite{Fye,Assaad94b}.
As we will see, this condition is satisfied in  our calculations 
and the systematic error produced by the Trotter decomposition is of order 
$ \Delta \tau ^2$.

Having isolated the two-body interaction terms, one may carry out 
Hubbard Stratonovitch transformations so as to express  the right hand
side of equation (\ref{Trotter})
as an imaginary time propagation of non-interacting electrons 
in an external field. 
For the Hubbard interaction, it is convenient to carry
out a discrete Hubbard Stratonovitch decomposition \cite{Hirsch83} to
obtain:
\begin{equation}
\label{HS}
      e^{-\Delta \tau H_U}  = C \sum_{\vec{s}}  \exp
      \left( \sum_{ \vec{i},\vec{j}, \sigma } c_{\vec{i}, \sigma }^{\dagger} 
      D_{\vec{i},\vec{j}}^{\sigma}(\vec{s}) c_{\vec{j}, \sigma} \right).
\end{equation}
Here $\vec{s}$ denotes a vector of length given by the number of sites,
$N$, of HS Ising fields and
\begin{equation}
D_{\vec{i}, \vec{j}}^{\sigma} (\vec{s}) =
\delta_{\vec{i},\vec{j}} 
{\rm cosh}^{-1} (\Delta \tau U/2) s_{\vec{i}} \sigma.
\end{equation}
The constant $C = {\rm exp} ( -\Delta \tau N  U/2)/ 2^N $ for the
$N$-site system will be dropped below. 

For the decomposition of $ H_{W}^{(n)} $ we have used the approximate 
identity:
\begin{equation}
\label{HSW}
	e^{-\Delta \tau H_W^{(n)} } =   C'
    \sum_{\vec{l}^{(n)} } a^{(n)}(\vec{l})
\exp  \left( \sum_{ \vec{i},\vec{j}, \sigma } c_{\vec{i}, \sigma}^{\dagger}
      A_{\vec{i},\vec{j}}^{(n)}(\vec{l}) c_{\vec{j}, \sigma} \right) + 
O(\Delta \tau ^4)
\end{equation}
where, $\vec{l}^{(n)}  = ( l^{(n,1)}, \dots, l^{(n,N/n_w)}) $,
$l^{(n,r)}  = -2, -1, 1, 2 $ for $r = 1 \dots N/n_w $ and
\begin{eqnarray}
& &  A_{\vec{i},\vec{j}}^{(n)}(\vec{l}) = \sqrt{ \Delta \tau W } 
\sum_{r=1}^{N/n_w} \eta \left( l^{(n,r)} \right) 
            A_{\vec{i},\vec{j}}^{(n,r)}  \nonumber \\
& &  a^{(n)}(\vec{l}) = \prod_{r = 1}^{N/n_w} \gamma \left( l^{(n,r)} \right)
          \nonumber \\
A_{\vec{i},\vec{j}}^{(n,r)} = & & 
	\delta_{\vec{i}, \vec{k}(n,r)}  
\left(  \delta_{\vec{j},\vec{i} + \vec{a}_x } e^{-i \phi} +
        \delta_{\vec{j},\vec{i} - \vec{a}_x } e^{ i \phi} +
        \delta_{\vec{j},\vec{i} + \vec{a}_y } +
        \delta_{\vec{j},\vec{i} - \vec{a}_y } \right)    \nonumber \\ +
& & \delta_{\vec{j}, \vec{k}(n,r)}
\left(  \delta_{\vec{i},\vec{j} + \vec{a}_x } e^{ i \phi} +
        \delta_{\vec{i},\vec{j} - \vec{a}_x } e^{-i \phi} +
        \delta_{\vec{i},\vec{j} + \vec{a}_y } +
        \delta_{\vec{i},\vec{j} - \vec{a}_y } \right). 
\end{eqnarray}
The fields $\eta$ and $\gamma$ take the values:
\begin{eqnarray}
 \gamma(\pm 1) = 1 + \sqrt{6}/3, \; \; \gamma(\pm 2) = 1 - \sqrt{6}/3
\nonumber \\
 \eta(\pm 1 ) = \pm \sqrt{2 \left(3 - \sqrt{6} \right)},  \; \;
 \eta(\pm 2 ) = \pm \sqrt{2 \left(3 + \sqrt{6} \right)} 
\nonumber  \\
\end{eqnarray}
The constant $C' = (1/4)^{N/n_w}$ will also be  dropped below.
The matrices $ A^{(n,r)} $ and $ A^{(n,r')} $ commute for all 
combinations of $r$ and $r'$.  The systematic
error produced by the above decomposition of  $H_W^{(n)}$ is of order 
$\Delta \tau ^4$ per time slice, and hence produces an overall systematic
error of order $\Delta \tau ^3 $.  Compared to the accuracy of the
Trotter decomposition (of order $\Delta \tau ^2$ ) this is negligible.

The trial wave function is required to be given by a Slater
determinant:
\begin{eqnarray}
\label{Trial}
     |\Psi_T \rangle  = & & |\Psi_T^{\uparrow} \rangle 
                \otimes |\Psi_T^{\downarrow} \rangle \nonumber \\
& & |\Psi_T^{\sigma} \rangle =  \prod_{n=1}^{N_p^{\sigma}} 
      \left( \sum_{\vec{i}} 
      c_{\vec{i},\sigma}^{\dagger} P_{\vec{i},n}^{\sigma} \right) |0
^{\sigma} \rangle .
\end{eqnarray}
Here $N_p^{\sigma}$ denotes the number of particles in a given spin
sector, $|0 ^{\sigma} \rangle$ is the vacuum in the spin-$\sigma$
sector and $P^{\sigma}$ is an 
$N \times N_p^{\sigma}$ rectangular matrix where $N$ is the number of sites.

One may now integrate the fermionic degrees of freedom and define
a  probability distribution: 
\begin{equation}
       {\rm Pr } (\vec{s},\vec{l} ) =  
\frac{ a(\vec{l} )
        \det \left(  M^{\uparrow} ( \vec{l}, \vec{s} )  \right)
        \det \left(  M^{\downarrow} ( \vec{l}, \vec{s} )  \right) }
    {\sum_{\vec{s}, \vec{l} }  a(\vec{l} ) 
	\det \left(  M^{\uparrow} ( \vec{l}, \vec{s} )  \right)
	\det \left(  M^{\downarrow} ( \vec{l}, \vec{s} )  \right) }. 
\end{equation}
The fields $\vec{s}$, $\vec{l}$ acquire an additional time index
$\tau : 1 \dots m $,  $a(\vec{l} ) = \prod_{r,n,\tau}
\gamma \left( l^{(n,r)}_{\tau}  \right) $  and
\begin{eqnarray}
    M^{\sigma} ( \vec{l}, \vec{s} ) & = &
         P^{\dagger, \sigma} B_{\vec{s}, \vec{l} }^{\sigma}(2\Theta, 0 ) 
         P^{\sigma}
\nonumber \\
   & & B_{\vec{s}, \vec{l}} (\Theta_2, \Theta_1) = 
         \prod_{ \tau = \tau_1 + 1}^{ \tau_2}
          e^{D^{\sigma}_{\tau}(\vec{s})} e^{A^{(1)}_{\tau}(\vec{l})} 
     \dots   e^{A^{(n_w)}_{\tau}(\vec{l})}
          e^{-\Delta \tau T }.
\end{eqnarray}
Here, $ \Theta_1 = \tau_1 \Delta \tau $,  $ \Theta_2 = \tau_2 \Delta \tau$ 
and $ H_t = 
\sum_{ \vec{i},\vec{j}, \sigma } c_{\vec{i}, \sigma}^{\dagger}
      T_{\vec{i},\vec{j}} c_{\vec{j}, \sigma}  $ corresponds to the
kinetic energy of the Hubbard model.
Observables may now be evaluated by:
\begin{equation}
\label{PQMC}
  \frac{ \langle \Psi_T |e^{-\Theta H }
          O
         e^{-\Theta H } | \Psi_T \rangle }
       { \langle \Psi_T |e^{-2\Theta H } | \Psi_T \rangle }  
=  \sum_{\vec{s}, \vec{l} }  {\rm Pr } (\vec{s},\vec{l} ) 
\langle O \rangle_{  \vec{s},  \vec{l} }   \; \; +  O \left( \Delta \tau ^2
\right)
\end{equation}
For a given set of fields $\vec{s}$,  $\vec{l}$,  Wicks theorem applies
and it suffices to calculate single-particle Green functions. They
are given by:
\begin{equation}
 \langle  c_{\vec{i},\sigma} c_{\vec{j},\sigma'}^{\dagger} 
 \rangle_{  \vec{s},  \vec{l} }  = \delta_{\sigma, \sigma'} 
\left( I - B_{\vec{s}, \vec{l}}^{\sigma} (\Theta,0) P^{\sigma} 
    M^{\sigma}(\vec{s},\vec{l})^{-1} 
       P^{\dagger,\sigma} B_{\vec{s},\vec{l}}^{\sigma}
(2\Theta,\Theta) \right)_{\vec{i},\vec{j}}.
\end{equation}
Here $I$ is the unit matrix, $I_{\vec{i},\vec{j}} = \delta_{\vec{i},\vec{j}}$.

As mentioned previously, the Monte Carlo simulation, for the  half-filled 
case, does not suffer from the sign problem.   That 
$ {\rm Pr } (\vec{s}, \vec{l} ) $ is positive for all values of the fields
$\vec{s}$ and $\vec{l}$ follows by carrying
out a particle-hole transformation in say the up spin sector:
$ c^{\dagger}_{\vec{i},\uparrow } = (-1)^{i_x + i_y} 
h_{\vec{i},\uparrow } $. Since the electron vacuum 
$ |0^{\uparrow} \rangle $ is
given by $ |0^{\uparrow}  \rangle    =
\prod_{\vec{i}}  h^{\dagger}_{\vec{i}, \uparrow } 
|0^{h, \uparrow}  \rangle $ 
one has
\begin{equation}
\label{Sign1}
	| \Psi_T^{\uparrow} \rangle   = 
       \prod_{n=1}^{N_p^{\uparrow}} 
      \left( \sum_{\vec{i}} 
      h_{\vec{i},\uparrow} (-1)^{i_x + i_y} 
      P_{\vec{i},n}^{\uparrow} \right) \prod_{\vec{i}}
h^{\dagger}_{\vec{i},\uparrow } |0^{h, \uparrow} \rangle \equiv 
 \prod_{n=1}^{N - N_p^{\uparrow}} 
    \left( \sum_{\vec{i}} 
      h_{\vec{i},\uparrow}^{\dagger} \overline{\tilde{P}}_{\vec{i},n} \right) 
|0^{h, \uparrow} \rangle.
\end{equation}
The above equation defines $\tilde{P}$  and
$\overline{\tilde{P}}_{\vec{i},n}$ denotes the complex conjugate of 
$ \tilde{P}_{\vec{i},n}$. Under the above transformation,
one has:
\begin{equation}
\label{Sign2}
	 \det \left(M^{\uparrow}_{\vec{s}, \vec{l} }  \right)
 =  e^{ \alpha \sum_{\vec{i}, \tau}  s_{\vec{i},\tau} }
	    \overline { \det \left( \tilde{P} ^{\dagger}
B_{\vec{s},\vec{l} }^{\downarrow} (2 \Theta, 0 ) \tilde{P}  \right) }
\end{equation}
where $\alpha = {\rm cosh}^{-1} (\Delta \tau U/2)$.
The complex conjugation comes from the fact that if electrons feel a flux
$\Phi$, holes are submitted to a flux $ -\Phi$.  The complex conjugation 
changes the sign of the flux $\Phi$. 
Provided that the
particle number in given spin sectors satisfy 
$N - N_p^{\uparrow} = N_p^{\downarrow}$  and that the trial wave
function satisfies  $ \tilde{P}  = P^{\downarrow} $ one has:
\begin{equation}
\label{Sign3}
	\det \left(M^{\uparrow}_{\vec{s}, \vec{l} }  \right) 
= e^{ \alpha \sum_{\vec{i}, \tau}  s_{\vec{i},\tau} }
       \overline { \det \left(M^{\downarrow}_{\vec{s}, \vec{l} } \right) }
\end{equation}
from which the positivity of the probability distribution 
${\rm Pr} (\vec{s}, \vec{l})$ follows. We have used a trial wave function  
which satisfies the above
condition for the positivity of the probability distribution.  
We furthermore require $ | \Psi_T \rangle $ to be a spin singlet:
\begin{equation}
\label{Trial1}
    \sum_{\vec{i}, \vec{j}} 
     \vec{S}_{\vec{i}} \cdot  \vec{S}_{\vec{j}} | \Psi_T \rangle = 0
\end{equation}
where $ \vec{S}_{\vec{i}} $ is the spin operator on site $\vec{i}$.
An explicit construction of trial wave functions showing no sign problem,
being spin singlet and if necessary real representable in real space is
given in the appendix.

The sum over the Hubbard Stratonovitch fields is carried out with
Monte Carlo methods. One  sweeps sequentially through  space time
lattice and proposes single spin flip  updates. 
The ratio of new to old probabilities 
under a local change  $l^{(n,r)}_{\tau}  \rightarrow
l'^{(n,r)}_{\tau} $ is best calculated in a basis where the matrix 
$A^{(n,r)}_{\vec{i}, \vec{j}}  $ is diagonal. Since in this basis  
$A^{(n,r)}_{\vec{i}, \vec{j}}  $ has 
only two non-vanishing eigenvalues, the ratio  of probabilities 
involves the knowledge of four Green functions per spin sector. 
Note that under a proposed local move $s_{\vec{i}, \tau} \rightarrow
s'_{\vec{i},\tau} $ the ratio of probabilities is given by a single
Green function per spin sector. 
In case of acceptance of the spin flip, the required updating  is two 
times more expensive in the case of an $l^{(n,r)}_{\tau}  \rightarrow
l'^{(n,r)}_{\tau} $  move than for an $s_{\vec{i}, \tau} \rightarrow
s'_{\vec{i},\tau} $ move.
The numerical stabilization of the algorithm is identical to that
used for Hubbard calculations.  This concludes our  description of
the PQMC algorithm.

The finite temperature QMC algorithm, is a grand canonical
simulation which evaluates:
\begin{equation}
\label{FT}
  \langle O \rangle =  \frac {{\rm Tr} \left( e^{-\beta H } O \right) }
                             {{\rm Tr} \left( e^{-\beta H }   \right) }
\end{equation}
where the trace runs over the Fock space and $\beta$ denotes the inverse
temperature.  
The finite temperature QMC algorithm
is conceptually similar to that of the PQMC \cite{Loh}.
Both the PQMC \cite{Assaad96a} and finite temperature QMC algorithms 
may be used to calculate imaginary time displaced correlation functions.
The dynamical results presented here stem from the use of the Classic
Maximum Entropy method \cite{Jarrel,Linden} to analytically continue 
imaginary time data produced by the finite temperature QMC algorithm. 

\begin{figure}
\mbox{}\\[3.0cm]
\epsfbox{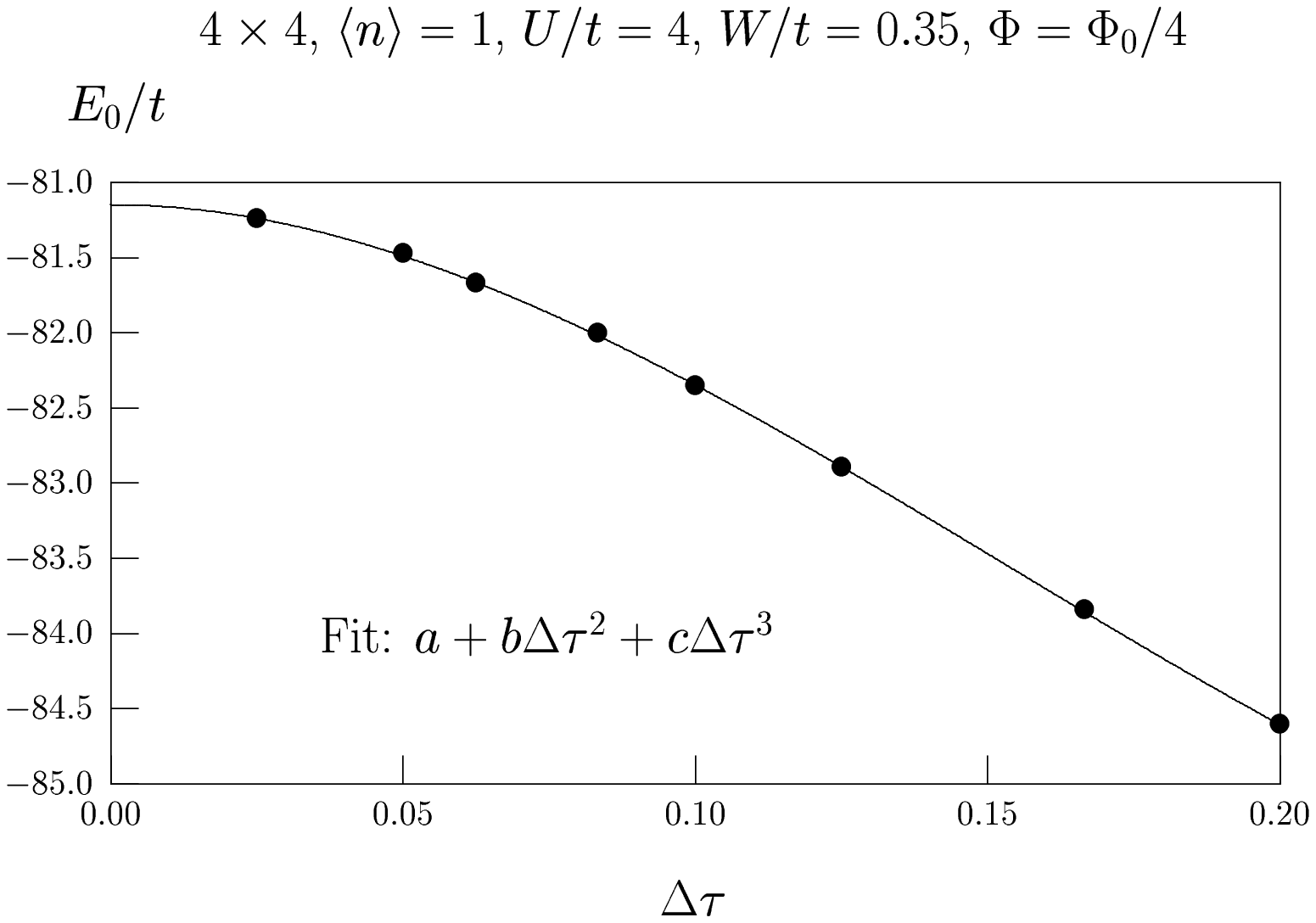}
\mbox{}\\[0.5cm]
\caption[]
{\noindent Total energy as obtained with the PQMC at $\Theta t = 1$ as a
function of $\Delta \tau$. The solid line is a least square fit to the
form $ a + b \Delta \tau^2 + c  \Delta \tau^3$ with 
$ a = -81.146 \pm 0.015 $,   $ b = -153 \pm 3 $ and $ c = 334 \pm 15 $. 
The energy unit is set by $ t = 1$.
\label{dtau.fig} }
\end{figure}

We are now in a position to test and compare the PQMC and finite 
temperature  algorithms. As mentioned previously, the systematic error
produced by the  Trotter decomposition is of the order $\Delta
\tau^2$  provided that the operators 
$H_W^{(n)}$, $H_t$, $H_U$, $O$, 
as well as the trial wave function $| \Psi_T \rangle $ are simultaneously real
representable. This may be checked explicitly by calculating the 
energy with the PQMC algorithm at $ \Phi/\Phi_0 = 0.25$, $ U/t = 4$,
$W/t =0.35$ on a $4 \times 4 $ lattice with  $2 \Theta t = 1$  for
various values of $\Delta \tau$. The results are plotted in Fig. 
\ref{dtau.fig}.
They follow well an  $ a + b \Delta \tau^2 + c  \Delta \tau^3$ form.
Note that the $\Delta \tau^3$  term, originates from the approximate Hubbard
Stratonovitch decomposition of $H_W$.  We have carried out most of
our PQMC simulations at $\Delta \tau = 0.0625 t$ which is sufficiently 
small for our purposes.

Fig. \ref{theta.fig}  shows plots of the total energy $E$ as well as 
the spin structure
factor at $\vec{Q} = (\pi, \pi)$, $ S(\vec{Q}) = \frac{4}{3}
\sum_{\vec{r}}  e^{i \vec{Q} \vec{r} } \langle \vec{S}_{\vec{r}} 
\vec{S}_{\vec{0}} \rangle $ as obtained with the PQMC and finite temperature
algorithms. For the finite temperature data, we set $ \beta  = 2
\Theta $ and measure the observables with equation (\ref{FT}). The same
observables are evaluated with equation (\ref{PQMC}) and the  above
described trial wave function.  We set $\Delta \tau = 0.1$ and in the
case of the PQMC algorithm measure observables on the ten  central 
time slices. For observables which do not commute with the Hamiltonian,
this yields an effective projection parameter 
$\Theta^{eff} = \Theta - 0.5t$. 
We consider a $6 \times 6$ lattice at $U/t =4$, $W/t = 0.35$ and
$ \Phi = 0$.
The trial wave function used here,
is constructed as shown in  Eq. (\ref{Trial2}) and the orthogonal 
transformation used  is obtained by diagonalizing the 
Hamiltonian of equation (\ref{Dimer}) 
at a numerically infinitesimal value of $\delta$. We have used this 
trial wave function for all our calculations at $\Phi = 0$.
In the limit of large $\Theta t$, both the finite temperature 
and PQMC results converge to the same  value within the error bars. We
observe that ground state properties  are obtained  more
efficiently within the PQMC approach.

\section{ Charge Degrees of Freedom}
\label{charge}
	In this section, we consider the charge degrees of
freedom at zero and finite temperatures. We start by showing that
at zero temperature, half-band filling, and constant value of the
Hubbard repulsion, $U/t = 4$,  $W/t$ triggers a 
transition between an insulator and a $d_{x^2 - y^2}$ superconductor.
The transition  is found to occur, at $W_c/t \sim 0.3 $.
We then consider the superconducting state at $W/t = 0.35$ and give numerical
evidence for the occurrence of a finite temperature Kosterlitz-Thouless
transition. The temperature dependence of the  one-electron
density of states, $N(\omega)$, is analyzed. 

\subsection{Zero temperature} 
An efficient and general method to distinguish between an
insulator and superconductor is to compute the  ground state energy
as a function of a twist $\Phi$ in the boundary condition along one
lattice direction: $ E_0(\Phi)$. In the case of an insulator, 
the wave function is localized and hence, an exponential decay of 
$ \Delta E_0(\Phi) \equiv E_0(\Phi) -
E_0(\Phi_0/2)$ as a function of lattice size is expected \cite{Kohn}.
In the spin density wave (SDW)
approximation for the half-filled Hubbard model, one obtains
$ \Delta E_0(\Phi)  = \alpha(\Phi) L \exp \left( -L/\xi \right)$ where
$\xi$  is the localization length of the wavefunction.
On the other hand, for a superconductor (i.e off-diagonal long-range
order (ODLRO)), $ \Delta E_0(\Phi) $
shows anomalous flux quantization: $ \Delta E_0(\Phi)$ is a
periodic function of $\Phi$ with period $ \Phi_{0}/2 $ and a
non vanishing energy barrier is to be found between the flux minima
\cite{Byers,Yang,Assaad93,Assaad94b} so that $\Delta E_0(\Phi_0/4)$ 
remains finite as $L \rightarrow \infty $.  In general, this functional 
form of $ \Delta E_0(\Phi)$ should occur only in the thermodynamic limit.

\begin{figure}
\epsfbox{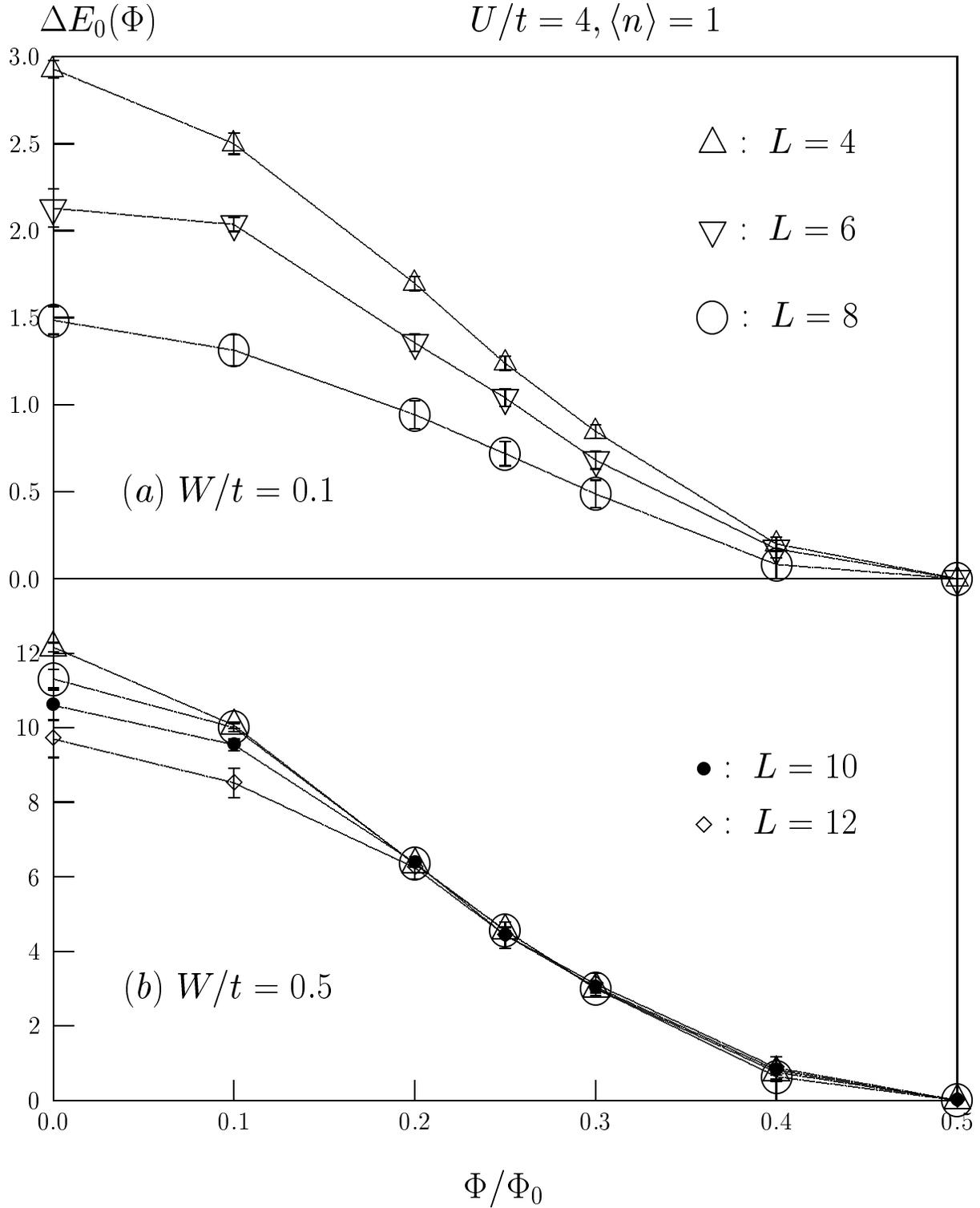}
\mbox{}\\[0.5cm]
\caption[]
{$ \Delta E_0(\Phi) \equiv E_0(\Phi) - E_0(\Phi_0/2)$ 
at (a) $W/t = 0.1$ (b) and $W/t = 0.5$ 
\label{flux_quant.fig} }
\end{figure}

Fig. \ref{flux_quant.fig} 
shows $\Delta E_0(\Phi) $ at $W/t =0.1$ and $W/t = 0.5$.
At $W/t =0.1$, and as expected for an insulator, $\Delta E_0(\Phi)$ 
is suppressed with growing lattice sizes for all values of $\Phi$.
In contrast, at $W/t = 0.5$ one notes that for values of 
$ 1/4 < \frac{\Phi}{\Phi_0} < 1/2 $, $ \Delta E_0(\Phi_0) $  is very 
stable against growing lattice sizes.
On the other hand, at  $\Phi = 0 $,  $ \Delta
E_0(\Phi) $ decreases with growing lattice sizes.  
We interpret the data
as a finite size indication of anomalous flux quantization.
At the end of Sec. (\ref{spinT0}) we will argue that this {\it slow}
size scaling of $ \Delta E_0(\Phi) $ 
in the vicinity of $\Phi = 0$ is related to the nature of the spin degrees 
of freedom in the superconducting state. 
Very similar results were obtained in the case of the repulsive and
attractive Hubbard models at half-filling \cite{Assaad93,Assaad94b}.

\begin{figure}
\epsfbox{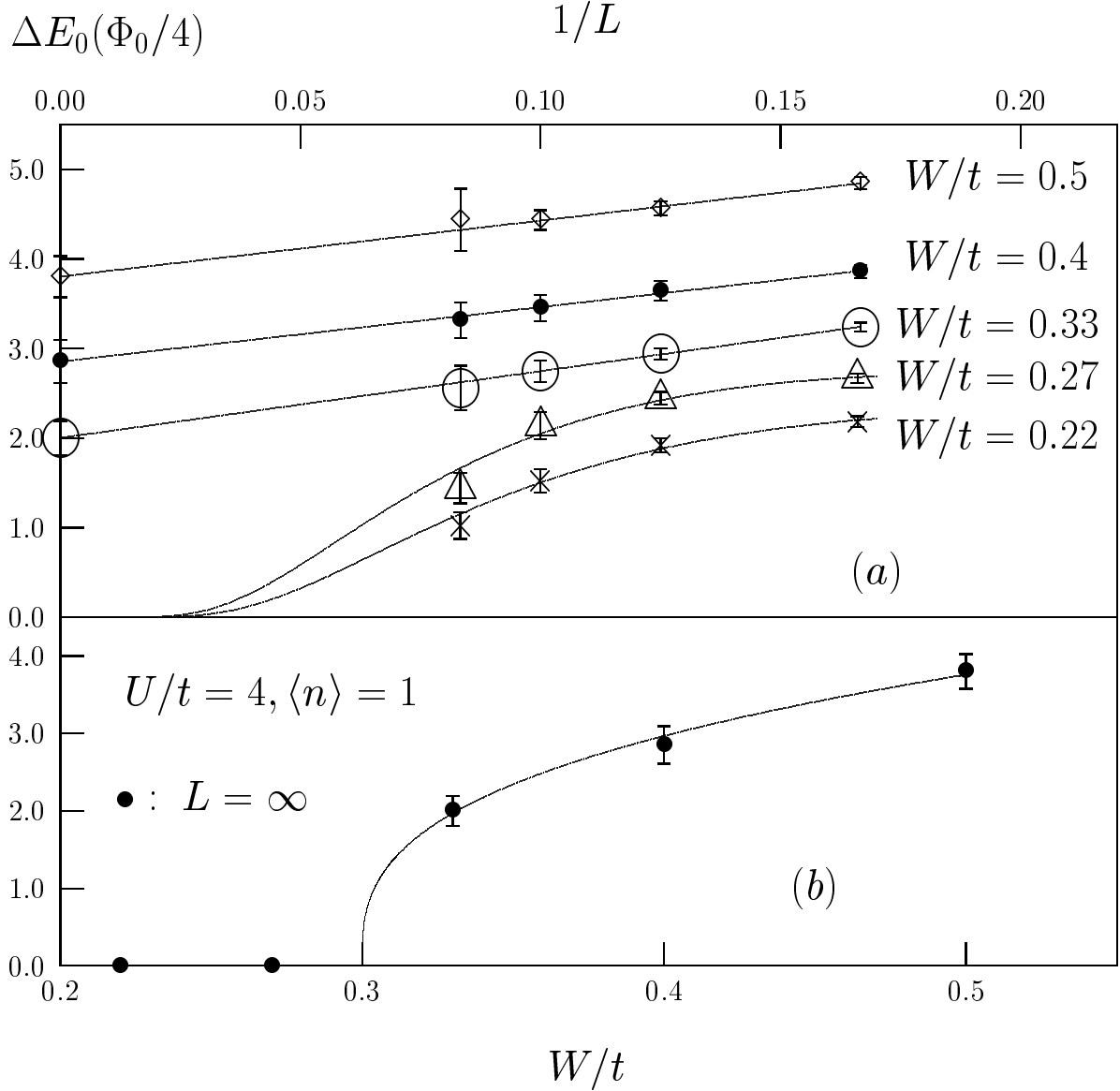}
\mbox{}\\[0.5cm]
\caption[]
{ (a) $\Delta E_0(\Phi_0/4) \equiv E_0(\Phi_0/4) - E_0(\Phi_0/2)$
versus $1/L$  for several values of $W/t$. For $W/t < 0.3 $  the solid
lines correspond to a least square fit of the data to the SDW form:
$L \exp(-L/\xi)$. For $W/t  > 0.3 $ the QMC data is compatible
with a $1/L$ scaling to a finite constant.
 The solid lines are least square fit to this form.
(b) Extrapolated value of $\Delta E_0(\Phi_0/4)$ versus $W/t$.
The solid line is a guide to the eye.
\label{su_den.fig} }
\end{figure}

To best distinguish between the insulating and superconducting states, 
we consider the size scaling of the quantity $\Delta E_0(\Phi_0/4)$ 
for various values of $W/t$ as in Fig. \ref{su_den.fig}a.
One observes a change in the size-scaling of $\Delta E_0( \Phi_0/4)$
as $W/t$ decreases from $W/t = 0.5$ to $W/t = 0.22$. From these
measurements, we estimate that the change occurs in the vicinity of
$W/t  = 0.3$.
For values of $W/t <   0.3$,  $\Delta E_0(\Phi_0/4)$ is consistent
with the SDW form  whereas for $W/t \geq  0.33$ $\Delta
E_0(\Phi_0/4)$ may be extrapolated to a  finite value with 
finite size corrections proportional to $1/L$.
The extrapolated value of $\Delta E_0(\Phi_0/4)$ versus $W/t$
is plotted in Fig. \ref{su_den.fig}b and the
quantum transition between a Mott insulator and superconductor occurs at
$W/t \sim 0.3 $.

\begin{figure}
\mbox{}\\[3.0cm]
\epsfbox{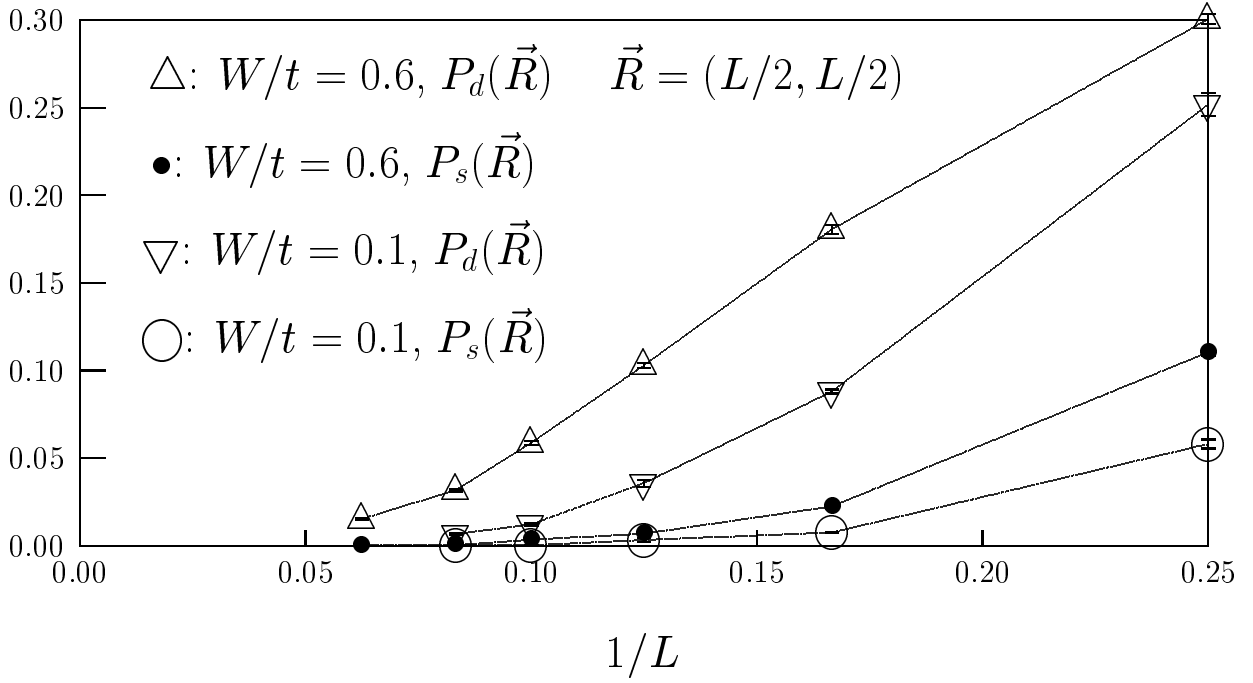}
\mbox{}\\[0.5cm]
\caption[]
{$d_{x^2 - y^2}$ (triangles) and $s$-wave (circles)
pair-field correlations  versus $1/L$. Those simulations were carried
out at $\Phi = 0$ (see Eq. (\ref{Bound})).
\label{PairT0.fig} }
\end{figure}

In order to determine the symmetry of the order parameter in the
superconducting state, we have calculated pair-field correlations
in the $s$ and $d_{x^2 - y^2}$ channels:
\begin{equation}
P_{d,s} (\vec{r}) = \langle \Delta_{d,s}^{\dagger}(\vec{r})
\Delta_{d,s}(\vec{0}) \rangle
\end{equation}
with
\begin{equation}
\Delta_{d,s}^{\dagger}(\vec{r})  =
\sum_{\sigma,\vec{\delta}}  f_{d,s}(\vec{\delta})
\sigma c^{\dagger}_{\vec{r},\sigma}
c^{\dagger}_{\vec{r} + \vec{\delta},-\sigma} .
\end{equation}
Here, $f_{s}(\vec{\delta}) = 1$ and $f_{d}(\vec{\delta}) = 1 (-1)$
for $\vec{\delta} = \pm \vec{a_x}$ ($\pm \vec{a_y})$.
Fig. \ref{PairT0.fig} shows plots of $P_{d,s} (L/2,L/2)$ for $W/t = 0.6$ where the
system
is superconducting and $W/t = 0.1$ where it is not.
At $W/t = 0.6$, the dominant signal
at long distances ($L = 16$) is obtained in the $d_{x^2 - y^2}$ channel.

\begin{figure}
\mbox{}\\[3.0cm]
\epsfbox{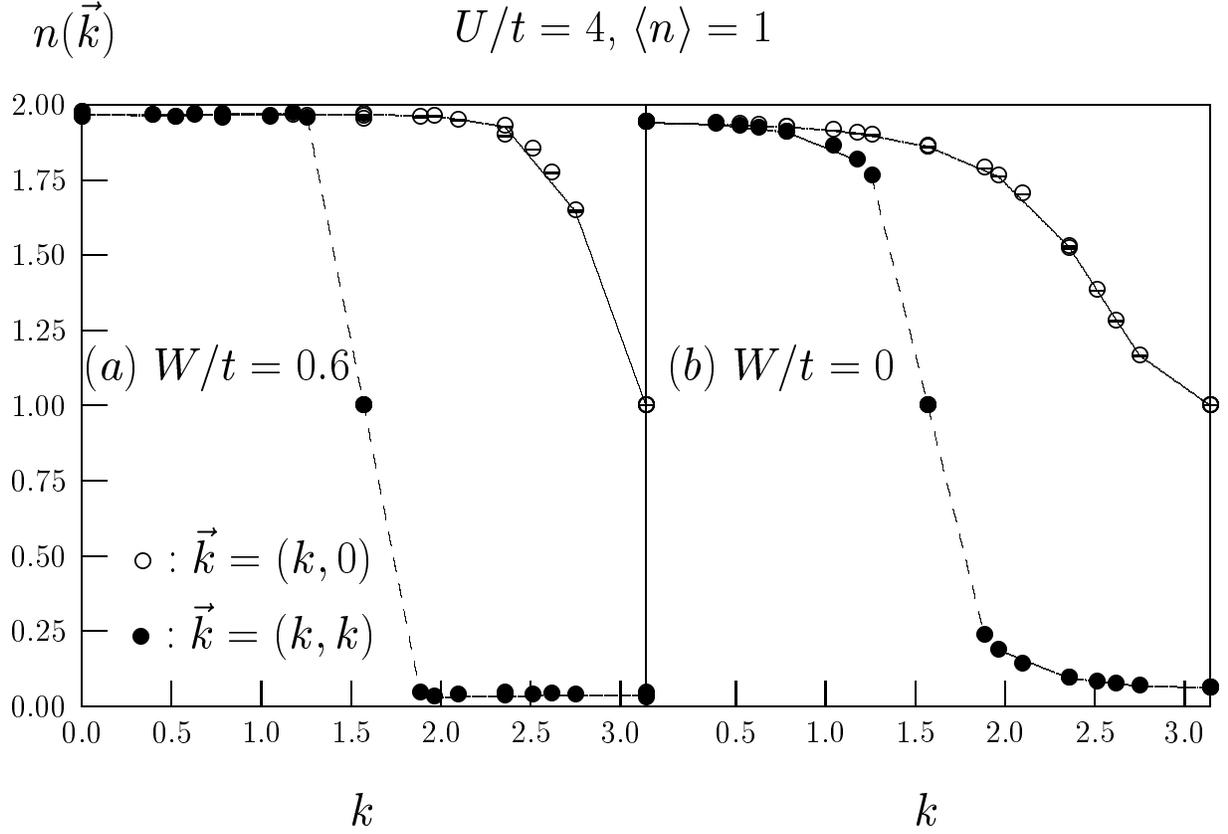}
\mbox{}\\[0.5cm]
\caption[]
{(a) $n(\vec{k})$ at $W/t = 0.6$, $U/t=4$ and $\langle n \rangle = 1 $.
Lattices form $ L=8 $ to $ L=16 $ were considered.
(b) same as (a) but for $W/t = 0$.  The calculations in this figure
were carried out at $\Phi = 0$ (see Eq. (\ref{Bound})).
\label{n_k.fig} }
\end{figure}

At the mean-field level, the symmetry of the order parameter will
determine the functional form of the single-particle occupation number,
$n(\vec{k})$. For a $d_{x^2 - y^2}$ superconductor the BCS result yields:
\begin{equation}
\label{nk}
n(\vec{k}) = 1 - \frac{\epsilon_{\vec{k}}}{
\sqrt{ \Delta_{\vec{k}}^2 +  \epsilon_{\vec{k}}^2 } }
\end{equation}
where $\epsilon_{\vec{k}}$ corresponds to the single-particle
dispersion relation and 
$ \Delta_{\vec{k}} = \frac{\Delta_0}{2} \left(\cos(k_x) - \cos(k_y) \right)$.
As is apparent from Eq. (\ref{nk}),  in the $ \vec{k} = k(1,1)$ direction
where the $d_{x^2 - y^2}$ gap vanishes 
$ n(\vec{k}) $ shows a jump at the Fermi energy, whereas in the
$ \vec{k} = k(1,0) $ direction  $ n(\vec{k}) $ is a smooth function of
$ k$.  Precisely this behavior in $ n(\vec{k})$ may be detected in the
QMC data at $W/t = 0.6$ as shown in Fig.\ref{n_k.fig}a. 
For comparison, we have plotted $n(\vec{k})$ at $W =0$ where it is 
expected to scale to a smooth function in the thermodynamic limit 
(see Fig.\ref{n_k.fig}b).

\subsection{Finite temperature: $W/t = 0.35$ } 
For values of $W/t > W_c/t \sim 0.3$, one expects the occurrence of a  finite
temperature Kosterlitz-Thouless transition. To detect this  transition 
we consider the quantity:
\begin{equation}
\label{DeltaET} 
   \Delta E \left( \Phi, T \right)  = E \left( \Phi = \Phi_0/4,T  \right) -
          E \left( \Phi = \Phi_0/2,T  \right).
\end{equation}
In the framework of a Kosterlitz-Thouless \cite{KT,KT1} transition one 
expects in the thermodynamic limit 
\begin{equation}
\label{KT}
 \Delta E \left( \Phi, T \right)   \sim \delta ( T - T_{KT} )
\end{equation}
where $T_{KT}$ is the Kosterlitz-Thouless transition temperature. 
The above follows by considering the Free energy  
$ F(\Phi ) =- \frac{1}{\beta}
\ln {\rm Tr} e^{-\beta H(\Phi)} $.
Expanding  $F$ around $ \Phi = \Phi_0/2 $ yields:
\begin{eqnarray}
F(\Phi ) & = & 
     F(\Phi_0/2 ) + \frac{1}{2}\left(\frac{\Phi -\Phi_0/2}{\Phi_0}\right)^2
                D_s(\beta) +
    O\left(\left(\frac{ \Phi -\Phi_0/2 }{\Phi_0}\right)^4\right) 
         \; \; {\rm where } \nonumber \\
& &  D_s(\beta)  = -  \left( \frac{2\pi}{L} \right)^{2}
 \left( <\tilde{K}_{x}> + \int_{0}^{\beta} d\tau <J_{x}(\tau)J_{x}(0)> \right),
     \nonumber \\
& &   J_{x}  =  \sum_{\vec{i}} t j_{\vec{i}}^{x}  + 
    W \left\{ K_{\vec{i}} + K_{\vec{i} + \vec{a}_x }, j_{\vec{i}}^{x} \right\} 
    \nonumber \\
& & \tilde{K}_{x} = 
          \sum_{\vec{i}} t k_{\vec{i}}^{x} - 2W \left( j_{\vec{i}}^{x} + 
          j_{\vec{i}-\vec{a}_x} \right)^2  + W \left\{ K_{\vec{i}} +
          K_{\vec{i} + \vec{a}_x }, j_{\vec{i}}^{x} \right\} 
     \nonumber \\
& & j_{\vec{i}}^{x}  = 
       -i \sum_{\sigma} 
     \left( c_{\vec{i},\sigma}^{\dagger} c_{\vec{i} + \vec{a}_{x}, \sigma } -
            c_{\vec{i} + \vec{a}_{x},\sigma}^{\dagger} c_{\vec{i}, \sigma}
     \right) \; \; {\rm and} \nonumber \\
& &  k_{\vec{i}}^{x}  =  - \sum_{\sigma}
\left( c_{\vec{i},\sigma}^{\dagger} c_{\vec{i} + \vec{a}_{x}, \sigma } \; + \;
       c_{\vec{i} + \vec{a}_{x},\sigma}^{\dagger} c_{\vec{i}, \sigma} \right).
\end{eqnarray}
The boundary conditions  for the calculation  of the superfluid density 
$ D_s$ satisfies: 
$c_{\vec{i} + L \vec{a}_x, \sigma } = - c_{\vec{i}, \sigma} $ and 
$c_{\vec{i} + L \vec{a}_y, \sigma } = c_{\vec{i}, \sigma}$. 
Hence,
\begin{equation}
\label{TKT}
	\Delta E \left( \Phi, T \right)  =  
         \frac{\partial} {\partial \beta } 
        \left( \beta F(\Phi) - \beta F(\Phi_0/2) \right)  
        \sim   \frac{1}{2} \left(\frac{ \Phi -\Phi_0/2 }{\Phi_0}\right)^2 
    \left( D_s - T \frac{ \partial } { \partial T } D_s  \right)
\end{equation}
At $T_{KT}$, $D_s$ is expected to show a universal jump \cite{KT1} so that
$ \frac{ \partial } { \partial T } D_s $ behaves like a  Dirac 
$\delta$-function and equation (\ref{KT}) follows.   Such signatures of 
the Kosterlitz-Thouless transition have already been observed in 
classical \cite{Himbergen} and quantum XY models 
\cite{Loh85} as well as in the attractive Hubbard model
\cite{Assaad94c,Assaad94b}.

\begin{figure}
\epsfbox{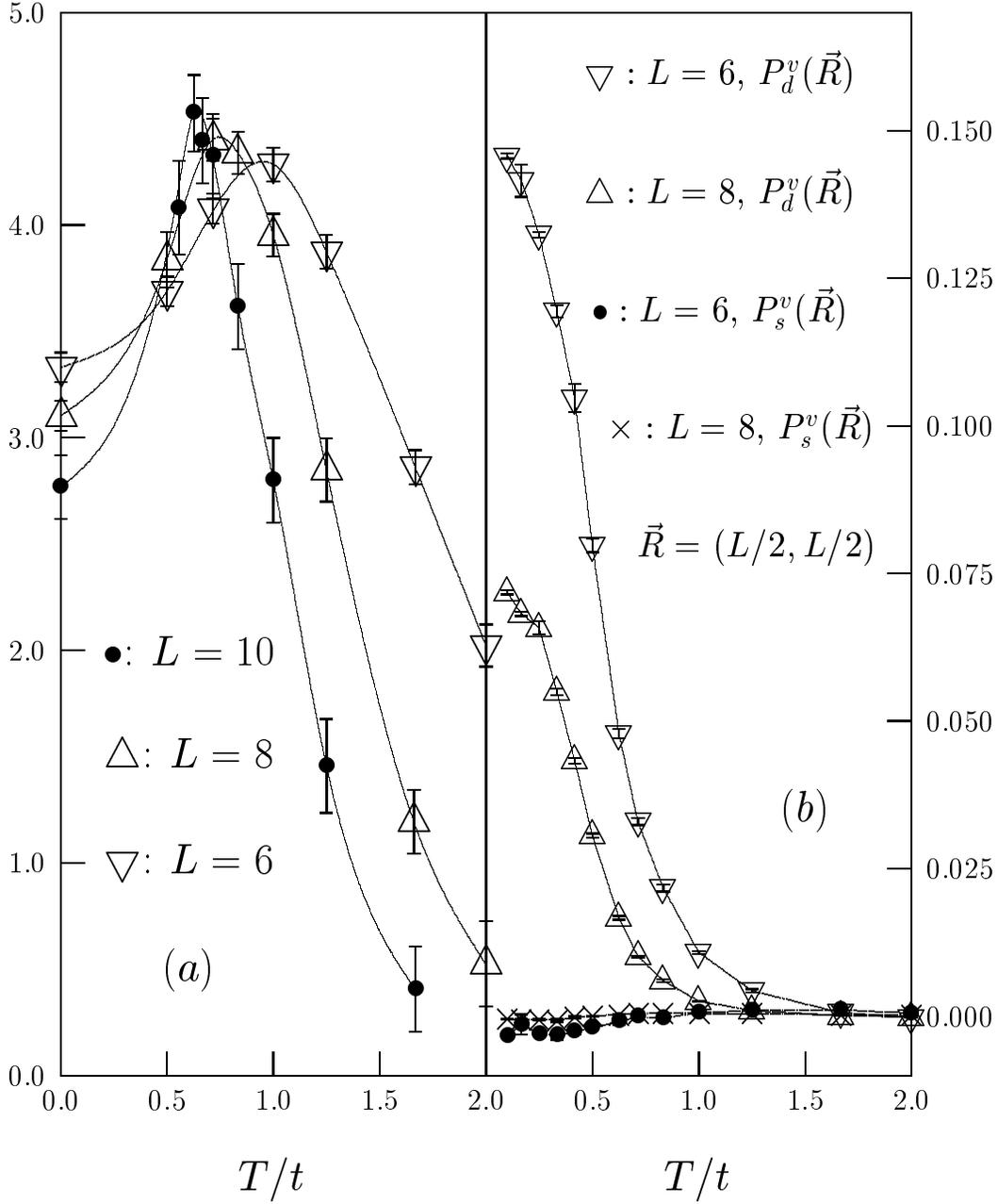}
\mbox{}\\[0.5cm]
\caption[]
{(a) $\Delta E (\Phi, T)$  as defined in equation (\ref{DeltaET}).
(b) Vertex contribution of the equal time pair-field
correlations  as defined in equation (\ref{Pair_vertex}).
Here we use periodic boundary conditions in both lattice directions
(i.e. $\Phi = 0$ in  Eq. (\ref{Bound})).

\label{tkt1.fig} }
\end{figure}

Figure  \ref{tkt1.fig}a  plots  $\Delta E (\Phi, T) $ at $W/t = 0.35$ and 
$U/t =  4$ for lattice sizes ranging from $ L = 6$  to $ L = 10 $. 
For each considered lattice size, $\Delta E (\Phi, T) $ shows a maximum 
at finite temperature.  For a given lattice size, we define $T_{KT}$ by 
this maximum and obtain:
$T_{KT} \sim 0.95, 0.75, 0.65 $  for the lattice sizes $L = 6, 8,10 $
respectively.  
We note that the scaling of the peak height is influenced by the size 
dependence of $T_{KT}$ due to the linear $T$ term in front of 
$\frac{ \partial } { \partial T } D_s $ (see equation  (\ref{TKT})).
The extrapolation of $T_{KT}$ to the 
thermodynamic limit is hard.  However, we recall that the size scaling of
the zero temperature energy difference  $ \Delta E_0 (\Phi_0/4) $ followed
well  $a + b/L$ form. Since $ \Delta E_0 (\Phi_0/4) $ is proportional to the
zero temperature superfluid density, it is plausible to assume that the size
dependence of $T_{KT}$ equally follows an  $a  + b/L$ form.  The above
three finite size values of $T_{KT}$ are consistent with this Ansatz, and we
obtain 
\begin{equation}
\label{TkT(L)}
	T_{KT} \sim 0.2t  + 4.5 t /L  {\rm \; \; at \; \;} W/t = 0.35.
\end{equation}

Figure \ref{tkt1.fig}b plots the vertex contribution to the equal
time pair-field 
correlations \cite{White89}
both in the $s$ and $d_{x^2 - y^2}$ channels for the $L = 6$ and  $L = 8$ lattices  
at the largest distance: $\vec{R} = (L/2,L/2) $. This quantity is given by:
\begin{equation}
\label{Pair_vertex}
 P_{d,s}^{v} (\vec{r}) = P_{d,s} (\vec{r})  - 
	\sum_{\sigma,\vec{\delta}, \vec{\delta}' }  f_{d,s}(\vec{\delta})
        f_{d,s}(\vec{\delta}')
 \left( \langle c^{\dagger}_{\vec{r},\sigma} c_{\vec{\delta}',\sigma}   \rangle 
        \langle c^{\dagger}_{\vec{r}+\vec{\delta},-\sigma} c_{\vec{0},-\sigma}  
        \rangle
 + \langle c^{\dagger}_{\vec{r},\sigma} c_{\vec{0},\sigma} \rangle
   \langle c^{\dagger}_{\vec{r}+\vec{\delta},-\sigma} 
           c_{\vec{\delta}',-\sigma} \rangle \right).
\end{equation}
Per definition, $ P_{d,s}^{v}  (\vec{r}) \equiv 0 $ in the absence of 
interactions.
An enhancement of the $d$-wave signal  at an energy scale set by $T_{KT}$ 
is observed.  In contrast, the $s$-wave response vanishes for all 
considered temperatures. Contrary to the $\Delta E (\Phi,T)$ data, 
those calculations were carried out at $\Phi = 0$.

\begin{figure}
\epsfbox{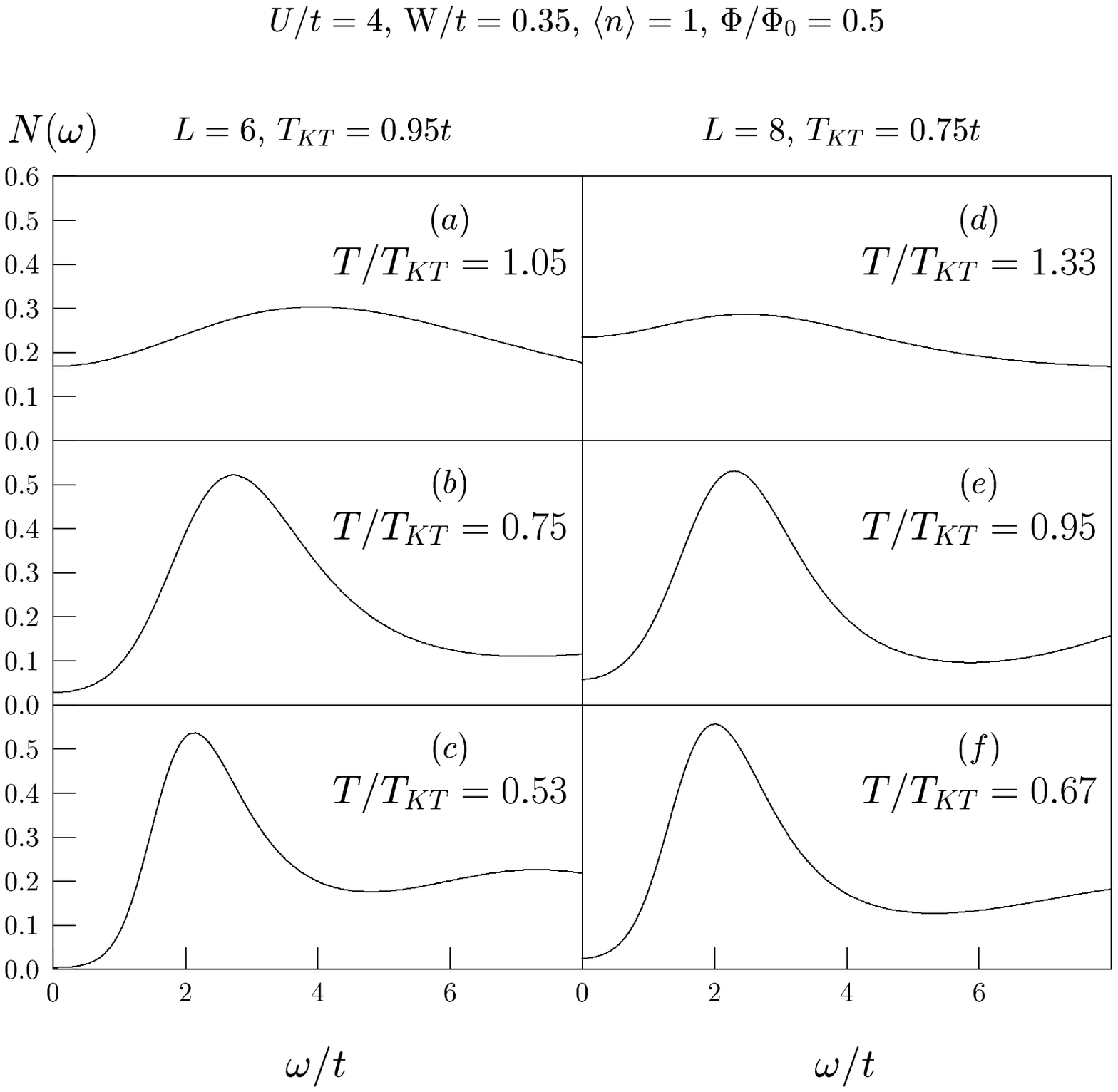}
\epsfbox{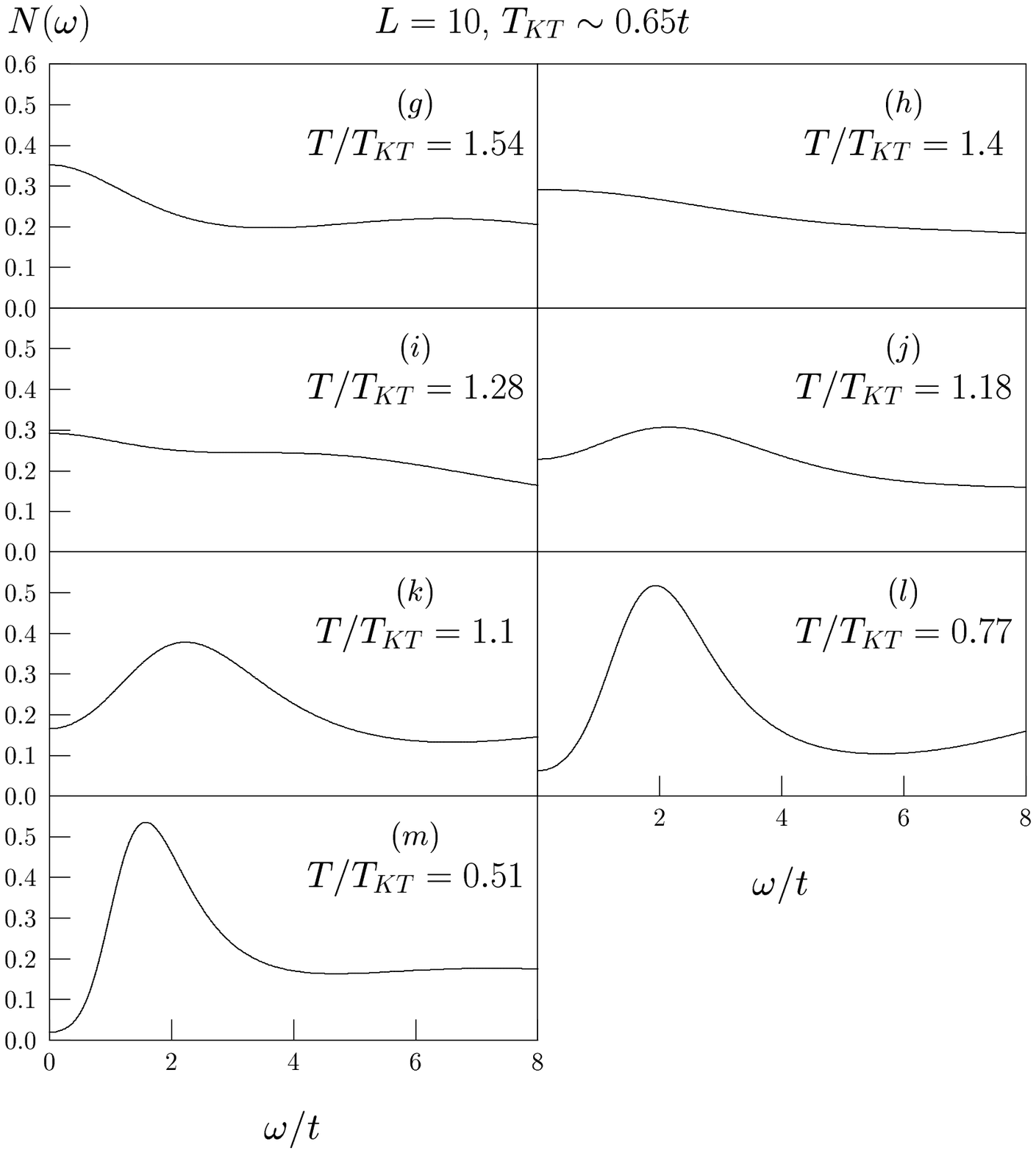}
\mbox{}\\[0.0cm]
\caption[]
{One-electron density of states as a function of temperature
and lattice size in the superconducting state at $W/t = 0.35$ and $U/t = 4$.
$ L = 6 $: Figs. $ (a)-(c) $. $ L = 8 $: Figs. $ (d)-(f) $.
$ L = 10 $: Figs. $ (g)-(m) $
\label{NOM.fig} }
\end{figure}

We now consider the  one-electron density of states:
$ N (\omega) = \frac{1}{N} \sum_{\vec{k}}  A(\vec{k}, \omega ) $ with 
\begin{equation}
      G(\vec{k},\tau) = \frac{1}{\pi} \int_{-\infty}^{\infty} {\rm d} \omega 
     \frac{e^{-\tau \omega}} { 1 + e^{-\beta \omega} } A(\vec{k}, \omega ) 
\end{equation} 
where $G(\vec{k},\tau) =   \sum_{\sigma} \langle  c_{\vec{k},\sigma} (\tau)
c^{\dagger}_{\vec{k},\sigma}  \rangle $, $\tau > 0$ and 
$ c_{\vec{k},\sigma} (\tau) = e^{\tau  H}  c_{\vec{k},\sigma} e^{-\tau  H}$.
We have used the Classic Maximum Entropy
method to obtain dynamical data from imaginary time Green functions.  We 
chose a flat default model for $N(\omega)$ and enforced the following
sum rules:
\begin{eqnarray}
  \int {\rm d} \omega N(\omega) & = & 2 \pi	 \nonumber \\
  \int {\rm d} \omega N(\omega) \omega {\rm tanh}  
        \left( \beta \omega / 2 \right)  & = & 
      \frac{\pi}{N} \sum_{\vec{k},\sigma}  
 \langle \left[  c_{\vec{k},\sigma}^{\dagger},  
 \left[H,c_{\vec{k},\sigma} \right]   \right] \rangle
\end{eqnarray}
The calculations were carried out at $\Phi = \Phi_0/2$.  The
data at temperatures lesser and greater than $T_{KT}$ is plotted
in Fig. \ref{NOM.fig}. We consider $ 6 \times 6 $ to $ 10 \times 10 $ 
lattices.
For all three considered lattices sizes, one sees
the onset of a pseudo-gap in the temperature range of 
$T_{KT}$. At {\it lower}  
temperatures, than shown in Fig. \ref{NOM.fig}, $N(\omega)$ suffers form size 
effects. 
The onset of finite size effects on small lattices (i.e. $L = 6$ and $L=8$ )
coincides approximately with the magnetic scale $J \sim 0.5 t$, which will be
determined in the next section. As the lattice size increases, one can go below 
the $J$ scale without noticing an anomaly in $ N(\omega)$. This may
be seen explicitly  in Fig. \ref{NOM.fig}(m) where $T = 0.33t < J $.
We may estimate the value of the superconducting gap $\Delta_{sc}$ by 
the peak position  in
$N(\omega)$, at the lowest temperature scale  presented in Fig. \ref{NOM.fig}.
For the three considered lattice sizes,
the data is consistent with  $\Delta_{sc}/ T_{KT} \sim 2.5$.
The peak position decreases as a function
of decreasing temperature. On the $10 \times 10$
lattice, we have looked more systematically at temperatures above $T_{KT}$.
The data shows a pseudo-gap feature above $T_{KT}$.
It is not clear if this aspect of the data will survive in the thermodynamic limit.

\section{Spin Degrees of Freedom}
\label{spin}
As in the previous section, we 
first concentrate on the zero temperature spin correlations and
show that the long-range antiferromagnetic order disappears at
$W_c/t  \sim 0.3 $. We then study finite temperature 
spin dynamics in the superconducting phase at $W/t = 0.35$.

\subsection{Zero temperature} 
\label{spinT0}
\begin{figure}
\mbox{}\\[-1.0cm]
\epsfbox{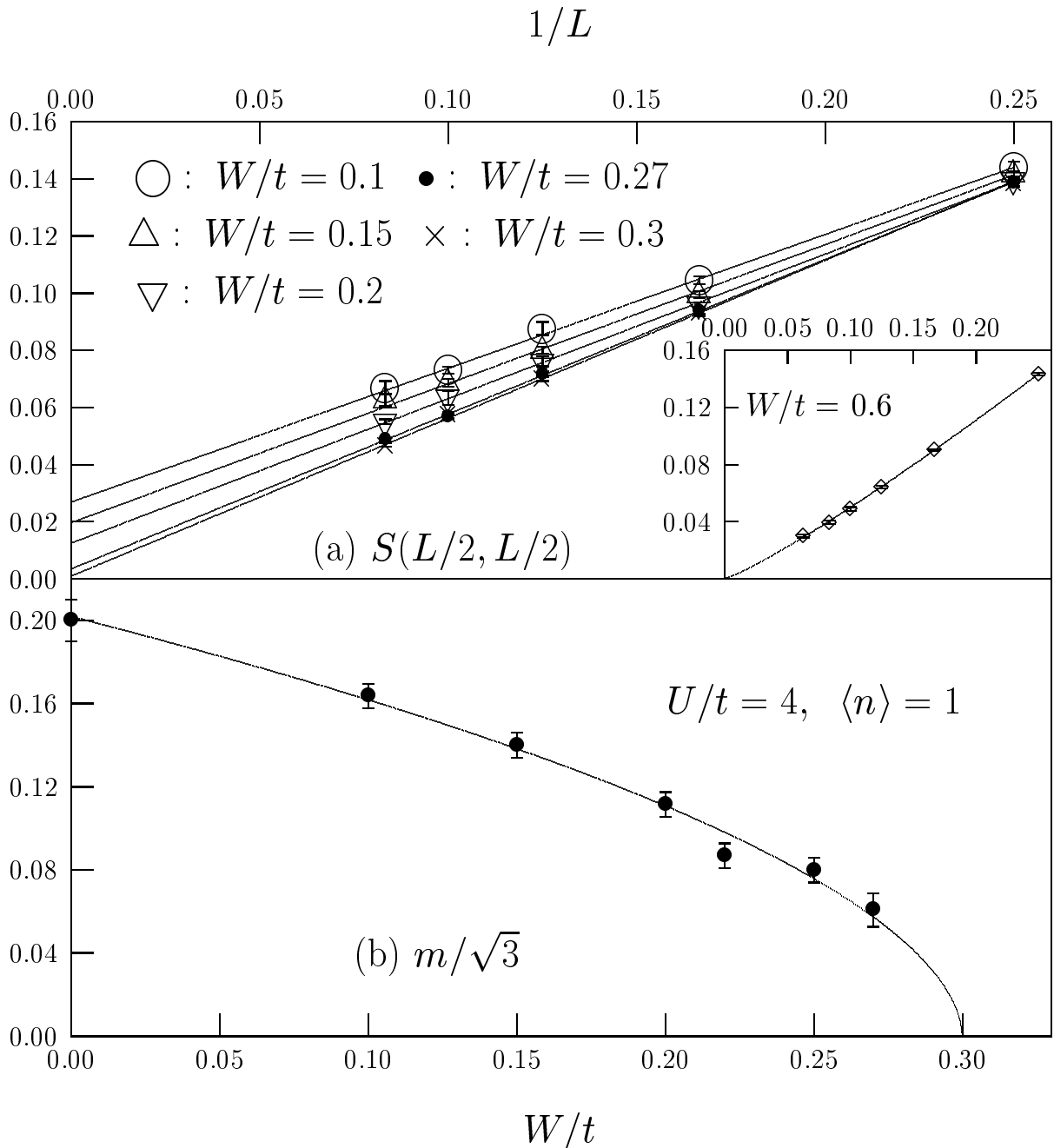}
\caption[]
{(a) $ S(L/2,L/2)$ versus $1/L$ for several values of $W/t$. The
solid lines correspond to least square fits of the QMC data to the form
$1/L$.  Inset: $ S(L/2,L/2)$ versus $1/L$  at $W/t = 0.6$. The solid
is a least square fit to the form $L^{-\alpha}$.
(b) Staggered moment  as obtained from (a) versus $W/t$.
The data point at $W/t = 0$ is taken from reference \cite{White}.
At $Wt/t = 0.3$, we were unable to distinguish $m$ from zero within our
statistical uncertainty.   The solid line is a guide to the eye.
The calculations in this figure
were carried out at $\Phi = 0$ (see Eq. (\ref{Bound})).
\label{spint0.fig} }
\end{figure}

To detect  the existence of long-range magnetic order,  we  compute the real
space spin-spin correlations:
\begin{equation}
S(\vec{r}) = \frac{4}{3} \langle \vec{S} (\vec{r} ) \vec{S} (\vec{0} ) \rangle
\end{equation}
where $ \vec{S} (\vec{r} ) $ denotes the spin operator on site
$\vec{r}$.
For values of $W/t < 0.3 $ and lattice sizes ranging from
$L=4$ to $L=12$,  $S(L/2,L/2) $,  may be fitted to a
$1/L$ form  and scales to a finite value, as shown in Fig. \ref{spint0.fig}a
We therefore conclude that long-range
antiferromagnetic order is present for $W/t < 0.3$.
The associated staggered moment, $ m = \lim_{L \rightarrow \infty}
\sqrt{3 S(L/2,L/2) } $,
is plotted in Fig. \ref{spint0.fig}b.  
The data is consistent with a continuous decay
of $m$ as $W/t$ increases towards $0.3$.
At $W/t = 0.3$, we were unable to distinguish
$m$ from zero within our statistical uncertainty. Hence, we conclude
that long-range antiferromagnetic order vanishes at $W/t \sim 0.3 $.
Therefore, within our numerical resolution, the antiferromagnetic
transition point is not separated from the superconductor-insulator
transition point.
Well within the $d_{x^2 - y^2}$ superconducting phase the
spin-spin  correlations remain sizable. In fact, at $W/t = 0.6$,
lattice sizes ranging from $L=4$ to $L=16$,  $S(L/2,L/2 ) $ scales
as $L^{-\alpha}$ with $\alpha = 1.16 \pm 0.01 $ as shown in the
inset of Fig. \ref{spint0.fig}a. 
This power-law decay of the spin-spin correlations in the
superconducting state arise because of the nodes in the $d_{x^2 - y^2}$
gap.
Following Ref. \cite{Nejat}, one can approximate the spin
susceptibility, $\chi(\vec{q}, i\omega_m)$, in the superconducting state
by inserting the irreducible
BCS spin susceptibility, $\chi_0^{BCS}(\vec{q}, i\omega_m)$, in the random
phase
approximation (RPA) form of  $\chi(\vec{q}, i \omega_m)$:
$\chi_{RPA}(\vec{q}, i \omega_m) = 
\chi_0^{BCS}(\vec{q}, i \omega_m)/ (1 - U
\chi_0^{BCS}(\vec{q}, i \omega_m) )$. Here, $\omega_m$ corresponds to
Matsubara frequencies.  Within this approximation and at half-band
filling,  the spin-spin correlations for a $d_{x^2 - y^2}$
superconducting order parameter show a power-law decay:
$S^{RPA} (L/2,L/2 ) \sim L^{-3.5}$. In contrast, the spin-spin 
correlations in an s-wave superconductor decay exponentially.  Thus, the
fact that spin-spin correlations remain critical in the superconducting
state,  may be attributed to the  node and phase factors of the
$d_{x^2 - y^2}$ superconducting gap. 

The very slow decay of the spin-spin correlations in the superconducting
state revealed by the QMC data  shows the extreme
compatibility  between a  $d_{x^2 - y^2}$ superconductor and substantial
short-range antiferromagnetic spin-spin correlations. In the superconducting
state, the spin stiffness vanishes since spin-spin correlations 
decay more rapidly than $\cos(\vec{Q}\vec{r})/|\vec{r}|$, where 
$\vec{Q} = (\pi,\pi)$.  However, since $S(\vec{r})$ decays
slower than $\cos(\vec{Q} \vec{r} ) /|\vec{r}|^2$ and the 
dimension is equal to two, 
$S(\vec{Q} ) \sim \sum_{\vec{r} } e^{i \vec{Q} \vec{r} }  
S(\vec{r}) $  diverges.  

At this point, we can offer an explanation of why it is convenient to 
study charge degrees of freedom  at $\Phi/ \Phi_0 = 0.5$ and 
spin degrees of freedom at $\Phi = 0$. We have just mentioned that in the 
superconducting state, the spin-spin correlations decay as 
$1/|\vec{r}|^{\alpha}$  where $ \alpha $ is slightly larger than one
( $ \alpha  \sim 1.16 $ at  $W/t = 0.6$). On a finite
lattice, it is hard to distinguish between true long-range order
and this slow decay. Choosing anti-periodic boundary conditions in
one lattice direction and periodic boundary condition in the
other (i.e. $\Phi/ \Phi_0 = 0.5$ ), renders the ground state of the
half-filled non-interacting system ($U = W = 0$) non-degenerate.
In a weak coupling approach, one may understand that this choice of
boundary conditions introduces frustration in antiferromagnetic 
correlations since on a finite lattice 
$\chi_0 ( \vec{Q}, \omega = 0 ) $ takes a finite value in the
zero temperature limit.  In contrast, and on a finite
lattice with periodic boundary conditions (i.e. $\Phi = 0$)
$\chi_0 ( \vec{Q}, \omega = 0 ) $ diverges in the zero temperature limit.
Here, $\chi_0$ denotes the
spin susceptibility at $ U = W = 0$. This frustration of the spin
degrees of freedom introduced by the boundary condition,
minimizes the size dependence of charge degrees of
freedom.  This  {\it pathology} may be seen explicitly in 
Fig. \ref{flux_quant.fig}b
where the superfluid density should in principle be given by  the
curvature of $\Delta E_0(\Phi) $ at $\Phi = 0 $ or $ \Phi =
\Phi_0/2$.  At $\Phi = 0$, the size scaling of this quantity is 
hard to extract from simulations up to $L=12$.  

\subsection{Finite temperature: $W/t = 0.35$ }

\begin{figure}
\mbox{}\\[-1.0cm]
\epsfbox{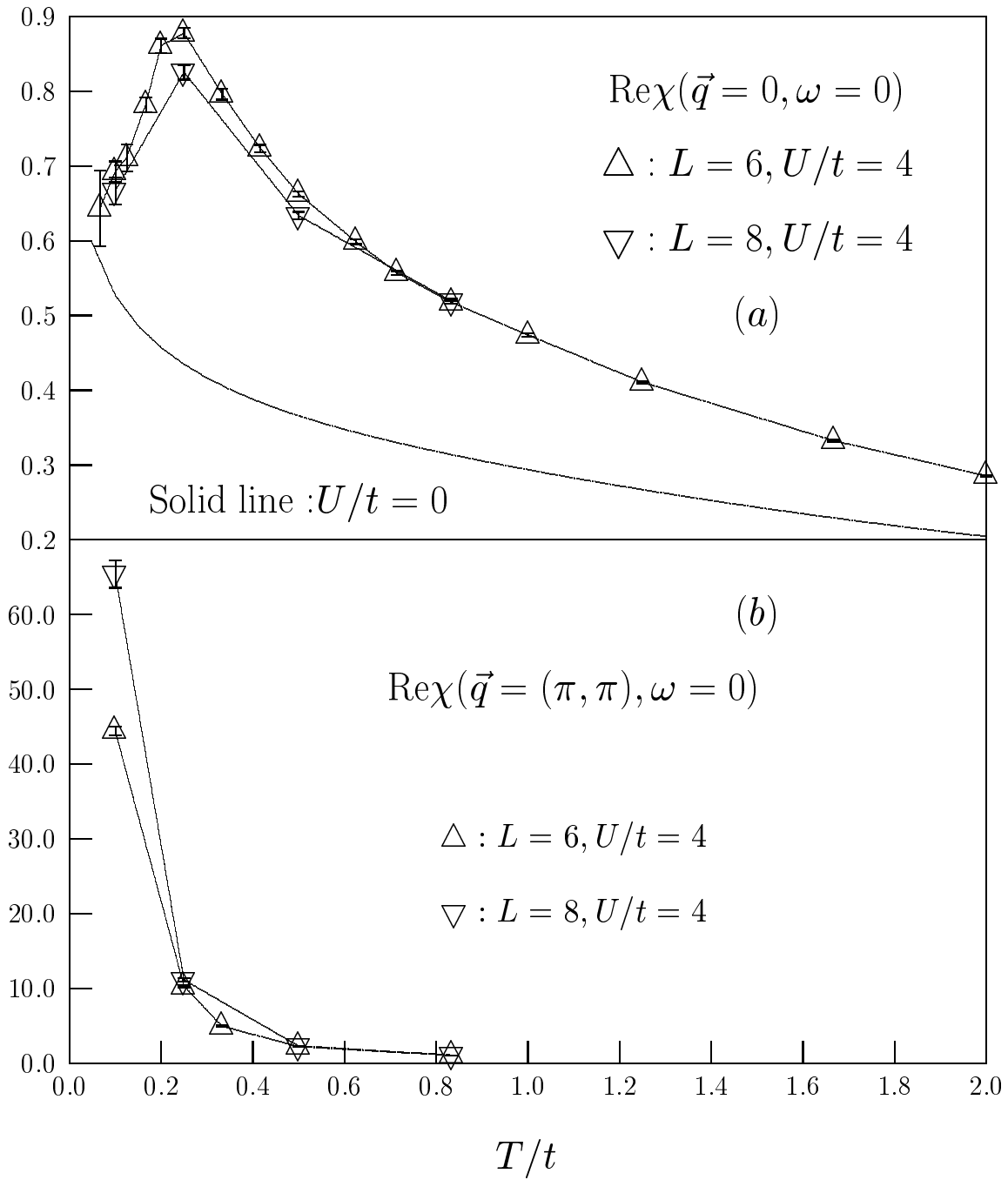}
\mbox{}\\[0.5cm]
\caption[]
{ (a) Uniform (b) staggered static spin susceptibility for the 
Hubbard model (i.e. $W/t = 0$) at $U/t = 4$.  The solid line with no 
symbols in (a) corresponds to 
$\chi (\vec{q}= 0, \omega = 0) $ at $U = W = 0$.
The calculations were carried out at $\Phi = 0$ (see Eq. (\ref{Bound})).
\label{ReChi0.fig} }
\end{figure}

We first consider the  static spin susceptibility,
\begin{eqnarray}
 {\rm Re } \chi \left( \vec{q}, \omega = 0 \right)  & \equiv &
 \int_{0}^{\beta} \langle m_z \left( \vec{q}, \tau \right)
                          m_z \left( -\vec{q},0 \right) \rangle  
\nonumber \\
   m_z \left( \vec{q},\tau \right)  =  
   e^{ \tau H } m_z \left( \vec{q} \right) e^{- \tau H}  \; \;& {\rm and} & \; \;
 m_z \left( \vec{q} \right) = \frac{1}{\sqrt{N}} \sum_{\vec{r}}
  e^{i \vec{q} \vec{r} }  \left( n_{\vec{r},\uparrow } -
n_{\vec{r},\downarrow } \right).
\end{eqnarray}
Before considering the superconducting phase, we briefly summarize the
temperature  dependence of  
${\rm Re } \chi \left( \vec{q}, \omega = 0 \right) $ at wave vectors 
$\vec{q} = 0$ and $ \vec{q} = \vec{Q} \equiv (\pi,\pi) $ for 
the Hubbard model at 
$U/t = 4$. The results are plotted in Fig. \ref{ReChi0.fig}. 
One expects the physics of the half-filled Hubbard model to be well described 
by a Heisenberg model in the  renormalized classical regime \cite{Chakravarty}.
Hence, an exponential
divergence   of $ {\rm Re} \chi( \vec{Q}, \omega =0 ) $  as a
function of temperature is expected.  The data 
(see Fig. \ref{ReChi0.fig}b) shows a sharp increase at a temperature 
scale $T \sim 0.25 $. We  use this criterion to
define the magnetic energy scale $J \sim 0.25 $ at $W/t = 0$. 
At temperatures $ T > J$ the uniform static susceptibility 
$ {\rm Re } \chi \left( \vec{q} = 0, \omega = 0 \right)  $ increases 
monotonously with decreasing temperature.  In this temperature regime, the
overall scale of $ {\rm Re } \chi \left( \vec{q} = 0, \omega = 0 \right) $, is
greater than its  free electron value.  This enhancement can be  understood
by the Stoner enhancement in the RPA approximation. In the low temperature
limit ${\rm Re } \chi \left( \vec{q} = 0, \omega = 0 \right) $  scales
to a finite value due to the existence of gapless spin-wave excitations. One 
notices that ${\rm Re } \chi \left( \vec{q} = 0, \omega = 0 \right) $ shows
a maximum at approximately $T \sim J$.  

\begin{figure}
\epsfbox{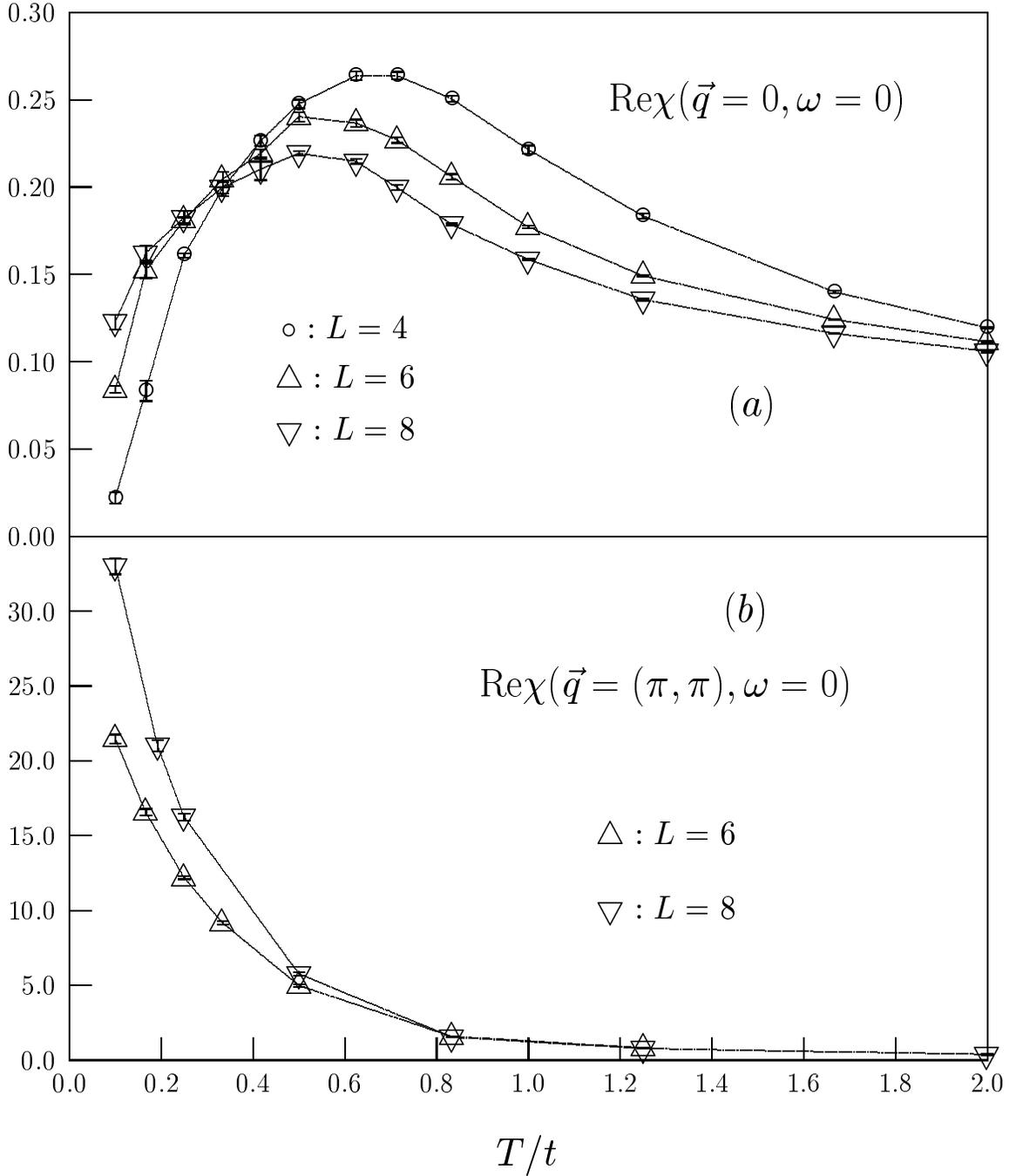}
\mbox{}\\[0.5cm]
\caption[]
{ (a) Uniform (b) staggered static spin susceptibility in the 
superconducting state at $W/t = 0.35$ and $U/t = 4$.  
The calculations were carried out at $\Phi = 0$ (see Eq. (\ref{Bound})).
\label{ReChi35.fig} }
\end{figure}

At $W/t = 0.35$ $ {\rm Re} \chi( \vec{Q}, \omega =0 ) $  
diverges in
the low temperature limit (see Fig. \ref{ReChi35.fig}b).
This divergence is weaker than at $W/t = 0$, 
and reflects the slow decay of the real space spin-spin correlations.
In the superconducting state the spin stiffness vanishes,
and one expects a power-law divergence of  $ {\rm Re} \chi(\vec{Q},
\omega =0 ) $  as a function of temperature.  From  this data, 
we may identify a magnetic energy scale $ J \sim 0.5 t$ at  $W/t = 0.35$.
Being a local quantity, $J$ is to first approximation insensitive to lattice
size.  That $J$ is greater at $W/t = 0.35$ than at $W/t =0$ may be seen
by inspecting the nature of the added interaction $H_{W}$. In the 
introduction, we have argued that at finite values of $U/t$, the 
relevant physics contained in the $H_{W}$ term  is given by $H_W^{(4)}$ 
in equation (\ref{HW1}).  As noted, $H_W^{(4)}$ contain terms of the form
$-W \Delta^{\dagger}_{\vec{i}, \delta} \Delta_{\vec{i}, \delta}$
which may be written as 
$ W  \left(  \vec{S}_{\vec{i}} \cdot \vec{S}_{\vec{i} + \vec{\delta} } -
\frac{1}{4} n_{\vec{i}} n_{\vec{i} + \vec{\delta} }   \right) $. 
This term explicitly enhances the value of $J$. 
At $W/t = 0.35$, and over the entire considered temperature region, 
$ {\rm Re } \chi \left( \vec{q} = 0, \omega = 0 \right)  $ 
is suppressed in comparison to its free  electron value 
(see Fig. \ref{ReChi35.fig}a). We interpret this overall suppression as a
signature of pairing fluctuations.    For our two largest lattice sizes,
$L = 6$ and $ L = 8$, $ {\rm Re } \chi \left( \vec{q} = 0, \omega = 0 \right)$
shows a maximum at $ T \sim 0.5t $ which coincides with our previously
determined $J$ scale. The fact that this maximum does not scale with 
system size (at least for our two largest lattices) makes us believe 
that it is related to the $J$ scale and not to $T_{KT}$.
In the low temperature limit, and as expected for
a superconducting state, the QMC data is consistent with the
vanishing of $ {\rm Re } \chi \left( \vec{q} = 0, \omega = 0 \right)  $. 
We note that in BCS theory and for a $d_{x^2 - y^2}$ order parameter,
one obtains: $ {\rm Re } \chi \left( \vec{q} = 0, \omega = 0 \right)  
\sim T $ in the low temperature limit. 

$T_{KT}$ is determined by using different boundary conditions 
(see equation (\ref{DeltaET}))
than $ {\rm Re } \chi \left( \vec{q} , \omega = 0 \right)  $,  which we
calculate at $ \Phi = 0$ (see equation (\ref{Bound})). Hence care 
has to be taken when comparing those two quantities. Strictly speaking,
one should extrapolate the two data sets to the thermodynamic limit
and only then compare. This is especially true in our case, since the
size effects are strong  in the determination of $T_{KT}$ (see equation
(\ref{TkT(L)})). We have seen that $T_{KT}$ scales to $0.2 t$ in the 
thermodynamic limit. At this  temperature scale, no evident anomaly appears
with growing lattice size in the  QMC data in Fig. \ref{ReChi35.fig}.

\begin{figure}
\epsfbox{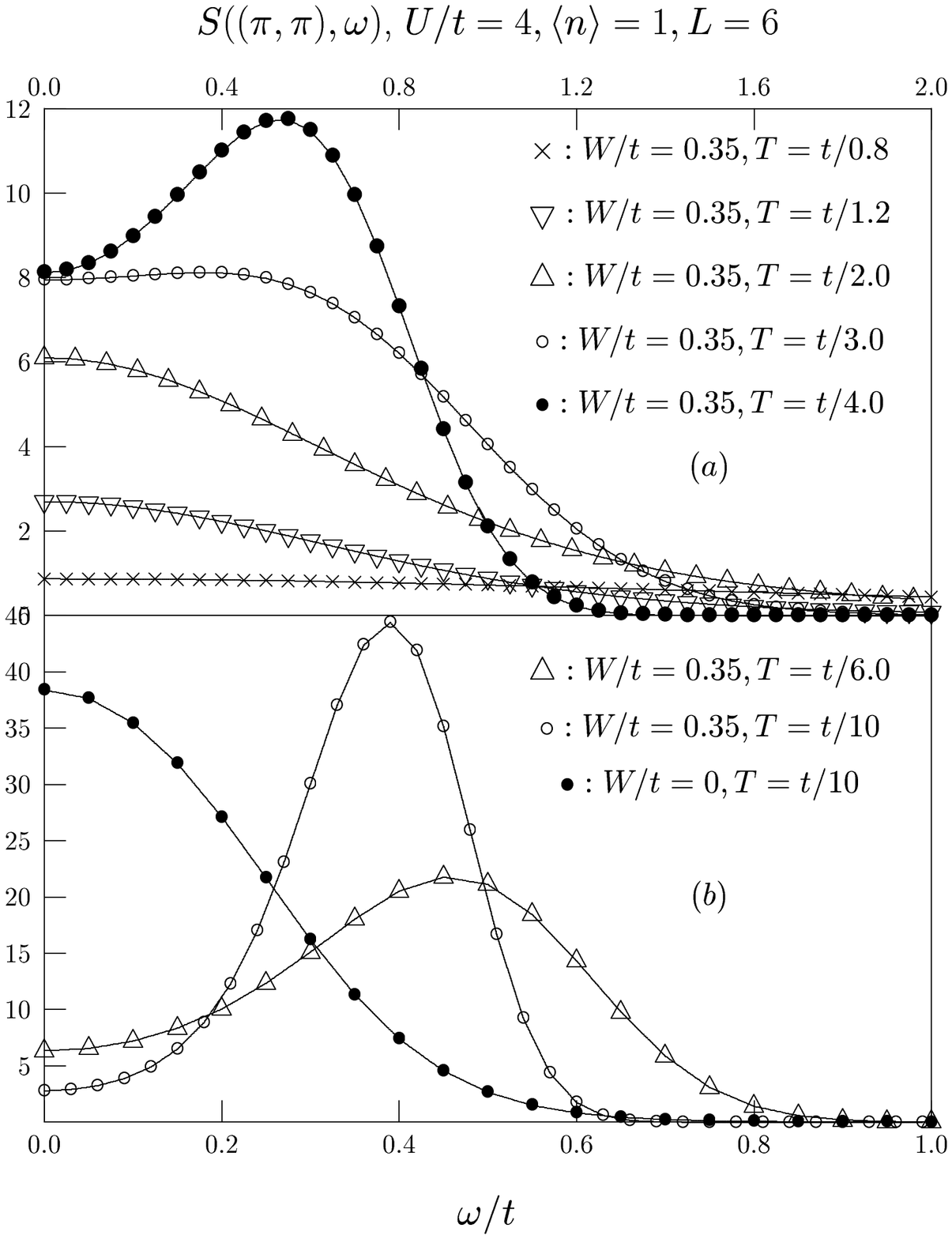}
\epsfbox{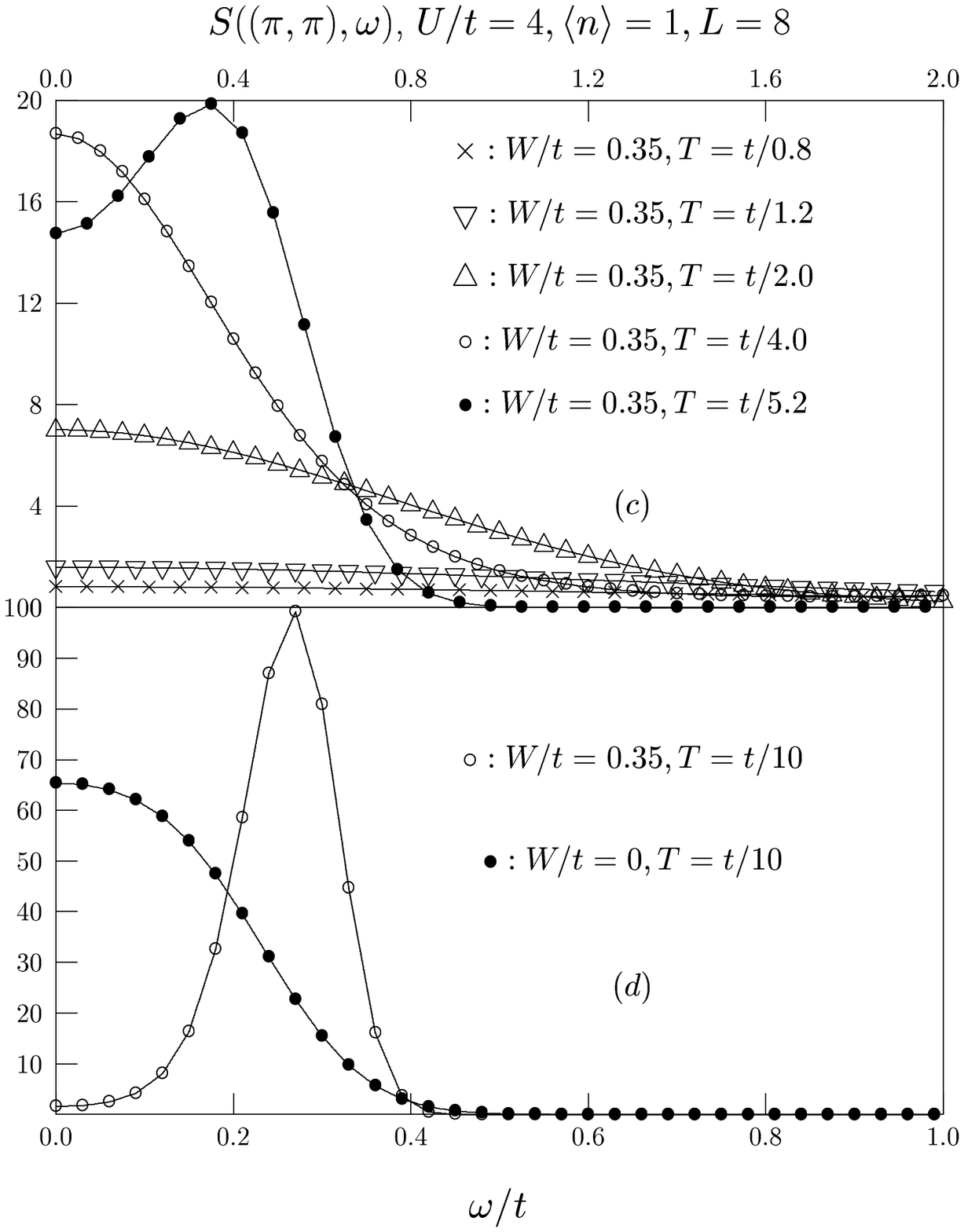}
\caption[]
{ $ S(\vec{Q}, \omega) $ in the superconducting state at $W/t =0.35$,
$U/t = 4$  as a function of system size and temperature.  For 
comparison, at the  lowest considered temperature,  $T = 0.1 t$, 
we have included the result for the Hubbard model (i.e. $W/t = 0$ ) 
at $U/t = 4$. 
Here, $\Phi = 0$ (see Eq. (\ref{Bound})).
\label{ImChi.fig} }
\end{figure}

We now consider the dynamical structure factor:
\begin{equation}
    S(\vec{q},\omega)  =    \frac{ {\rm Im } \chi (\vec{q}, \omega)  }
                                 { 1 - e^{-\beta \omega} }
\end{equation}
where 
\begin{equation}
\chi(\vec{q}, \omega ) =  -i \int_{0}^{\infty} {\rm d} t
e^{ i \left( \omega + i \delta \right)t } 
   \langle \left[ m_z \left( \vec{q},t \right),  m_z \left( -\vec{q},0 \right)
   \right] \rangle \; \; {\rm and} \; \;
   m_z \left( \vec{q},t \right)  =  
   e^{ i t H } m_z \left( \vec{q} \right) e^{-i t H}.
\end{equation} 
In the above, we have set $\hbar = 1$ and $\delta $ is a positive
infinitesimal number.   ${\rm Im } \chi (\vec{q}, \omega)$   was obtained 
by analytically continuing the imaginary time QMC data with the use of
the Classic Maximum Entropy method. We have used a flat default model for
$S(\vec{q}, \omega)$ and imposed the sum rules:
\begin{eqnarray}
 & &  \int_{-\infty}^{\infty} {\rm d} \omega 
         \frac{ {\rm Im } \chi (\vec{q}, \omega)  } {\omega}  = 
         \pi {\rm Re } \chi (\vec{q}, \omega = 0)  \nonumber \\
 & & \int_{-\infty}^{\infty}  {\rm d} \omega 
         {\rm Im } \chi(\vec{q}, \omega)  
    {\rm coth} \left( \beta \omega/ 2 \right)  = 
      2 \pi \langle m_z  \left( \vec{q} \right) m_z \left( - \vec{q} \right) 
          \rangle  \; \; {\rm and} \nonumber \\
 & & \int_{-\infty}^{\infty}  {\rm d} \omega
         {\rm Im } \chi(\vec{q}, \omega)  \omega = 
      \pi  \langle \left[  m_z \left( - \vec{q} \right) , 
           \left[  H, m_z \left( \vec{q} \right) \right] \right]
          \rangle.
\end{eqnarray}
Fig. \ref{ImChi.fig} plots $S(\vec{Q}, \omega) $ as a function of
temperature at $W/t = 0.35$  on $ 6 \times 6 $ and $ 8 \times 8 $
lattices. For comparison,  we have plotted at the lowest considered
temperature, $\beta  = 0.1t$, $S(\vec{Q}, \omega)$  for the Hubbard
model (i.e. $W/t = 0$).
$S(\vec{q}, \omega)$  satisfies the sum rule:
\begin{equation}
  \int_{-\infty}^{\infty}  {\rm d} \omega  S(\vec{q}, \omega ) = 
      \pi 
    \langle m_z \left( \vec{q} \right)  m_z \left( - \vec{q} \right) \rangle 
\equiv  \pi S(\vec{q}).
\end{equation}
At $W/t = 0$,  $S(\vec{Q})$ diverges as $L^2$ in the zero
temperature limit.  This weight is centered at $\omega = 0$.  
At $W/t = 0.35$,  and at a scale set by $J \sim 0.5 t$, we observe a 
buildup of spectral weight centered around $\omega = 0$. 
At a lower temperature scale, which we will identify below,  
spectral weight is shifted from lower frequencies to 
form a peak at finite frequency. Defining $\omega_0$ by
the peak value of $S(\vec{Q}, \omega)$  at the lowest considered temperature,
we see that $ \omega_0/T_{KT} \sim 0.4$. This relation is valid for
both considered lattice sizes. Since the equal time spin-spin 
correlations in the superconducting state decay slower than 
$\cos(\vec{Q}\vec{r})/|\vec{r}|^2 $ at zero temperature, the spectral
weight under the peak at $\omega_0$ diverges in the 
thermodynamic and low temperature limits. 

\begin{figure}
\mbox{}\\[-2.0cm]
\epsfbox{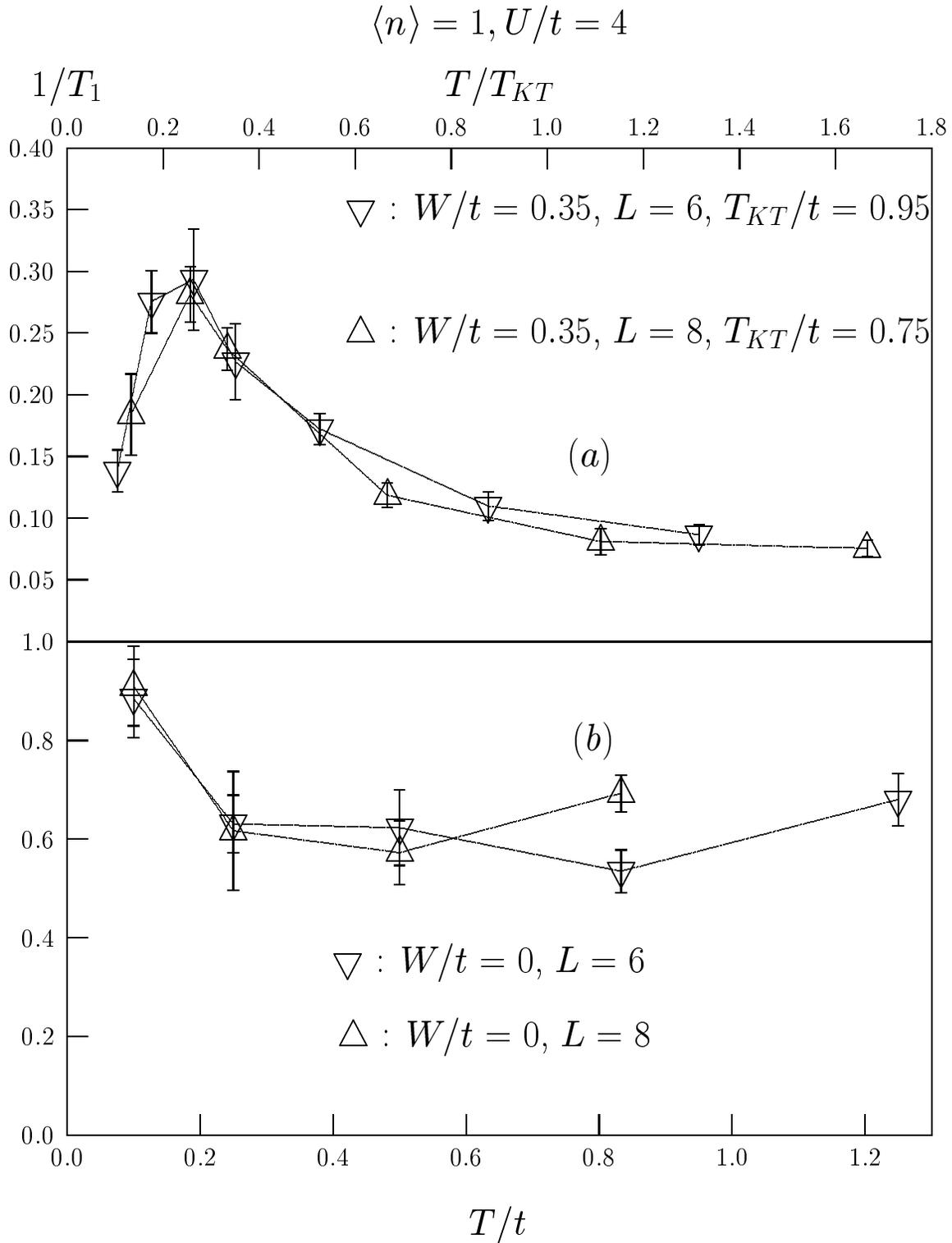}
\caption[]
{ (a) $1/T_1$ versus $T/T_{KT} $ in the superconducting state at
$W/t = 0.35$.  The magnetic scale is given by $J \sim 0.5 t$. In the
thermodynamic limit, $T_{KT} \sim 0.2 t$  and $J$ is to first approximation
size independent. Thus $J/T_{KT} \sim 2.5$ at $L \rightarrow \infty$. 
(b). $1/T_1$ versus $T$ for the  antiferromagnetic Mott
insulating state.  Here, $J \sim 0.25 t $.  
The calculations were carried out at $\Phi = 0$ (see Eq. (\ref{Bound})).
\label{T1.fig} }
\end{figure}

The relaxation rate we consider is defined by, 
\begin{equation}
   \frac{1}{T_1 T }  = \lim_{ \omega \rightarrow 0} 
           \frac{1}{N} \sum_{\vec{q}} 
         \frac{ {\rm Im} \chi \left( \vec{q}, \omega \right) }
               {\omega},
\end{equation}	
where a $ \vec{q} $-independent nuclear form factor is assumed.
Fig. \ref{T1.fig}  plots $1/T_1$ both in the superconducting and
antiferromagnetic insulating state.  The {\it error bars} are obtained
in the following way. For given imaginary time data and covariance
matrix, we  transform the data into a basis where the covariance matrix is
diagonal. In this basis, we add independent gaussian noise to each
data point. The width of each gaussian distribution is  determined
by the diagonal of the covariance matrix. We carry out the
analytical continuation and repeat the procedure for several
realization of the gaussian noise. The {\it error bars} in  Fig.
\ref{T1.fig} correspond to the variance of the so obtained values of
$1/T_1$. 
$1/T_1$ is a delicate quantity to compute with QMC methods. To check
the reliability of our calculations we first consider the $W/t = 0$ case.
At temperatures  $T < J$ as determined 
from ${\rm Re} \chi (\vec{Q}, \omega = 0)$, an increase 
in $1/T_1$ is observed. This is an  expected feature since in 
the renormalized classical regime of quantum
antiferromagnets,  $1/T_1$ diverges exponentially with decreasing
temperature \cite{Chakravarty}. 
In the superconducting state at $W/t = 0.35$, we plot
$1/T_1$ as a function of $T/T_{KT}$ (see Fig. \ref{T1.fig}a). 
The overall scale of $1/T_1$ is 
reduced in the superconducting state as compared to the
antiferromagnetic Mott insulating state. 
For both considered lattice sizes, $1/T_1$  shows a maximum at 
$T \sim 0.25 T_{KT}$ which allows us to define a cross-over temperature
scale: $ T^{cr}_{1}/T_{KT} \sim 0.25 $.   At temperatures lower than
$ T^{cr}_{1}$, the QMC data is consistent with the vanishing of the 
relaxation rate $1/T_1$. In BCS  theory with a 
$d_{x^2 - y^2}$ order parameter, $1/T_1 \sim T^3 $ in the low temperature 
limit.  
At temperatures above $T^{cr}_{1}$,  $1/T_1$ grows with 
decreasing temperature. One expects $1/T_1$ to start increasing  
at the magnetic  scale $J$ as determined from 
${\rm Re}  \chi(\vec{Q}, \omega = 0 )$.
Since in Fig. \ref{T1.fig}a we set the temperature scale by $T_{KT}$
large size effects are now present in $J/T_{KT} $ which takes the 
values $0.5$  ($0.65$) for the $L=6$ ($L=8$) lattice and  scales to 
$ J/T_{KT} \sim 2.5 $ in the thermodynamic limit.  
The so determined cross-over temperature coincides well with the temperature
scale at which spectral weight in $S(\vec{Q},\omega)$ is shifted
from low frequencies to form the peak centered at $\omega_0$ (see Fig. 
\ref{ImChi.fig}).

\section{Summary and Conclusions} 
\begin{figure}
\mbox{}\\[0.0cm]
\epsfbox{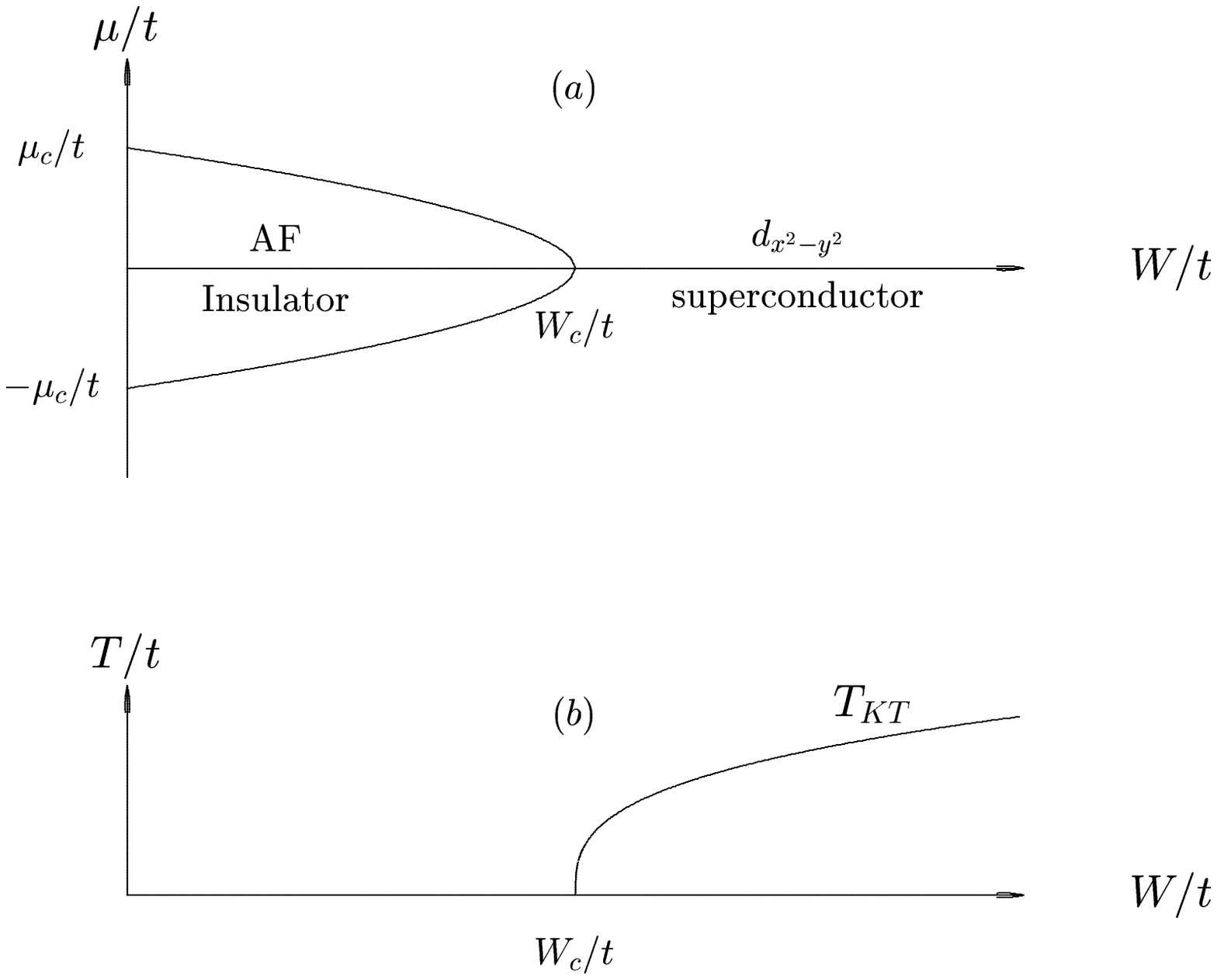}
\mbox{}\\[0.5cm]
\caption[]
{ (a) Schematic phase diagram  of the model $ H = H_U + H_W  - \mu \hat{N}$ 
at zero temperature and in the $W-\mu$ (see Eq. (\ref{tUW})). Here, $\mu$  
corresponds to the chemical potential and $\hat{N}$ is the particle number 
operator.  Half-filling corresponds to $\mu = 0$.  At $U/t = 4$ , $\mu = 0$,
we estimate $W_c/t \sim 0.3$ and $ \mu_c/t = 0.67 \pm 0.015$ \cite{Assaad96}.
(b) Schematic phase diagram  of $ H = H_U + H_W $ in the $T-W$ plane where
$T$ denotes the temperature. At $U/t = 4$  and $W/t = 0.35$ we estimate 
$T_{KT} \sim 0.2 t$.   \label{croc.fig}}
\end{figure}

As schematically illustrated in Fig. \ref{croc.fig}a our zero temperature 
results at $U/t = 4$  are best summarized in the   
$ W-\mu $ plane where $\mu$ denotes the chemical potential.
We have carried
out our simulations at half-filling, $\mu = 0$,  and shown the occurrence of
a quantum transition between an antiferromagnetic Mott insulator and 
$d_{x^2 - y^2}$ superconductor as a function of $W/t$.
Within our numerical accuracy, the quantum transition occurs simultaneously
in both the charge and spin degrees of freedom at $W_c/t \sim 0.3 $.  
Although the spin stiffness vanishes in the superconducting state, 
the antiferromagnetic spin correlations decay slower than 
$ 1/| \vec{r} |^2 $
where $|\vec{r}|$ denotes the distance. In two dimensions, this
slow decay leads to a divergence of the equal-time spin structure factor at
wave vector $ \vec{Q} = (\pi,\pi)$. This feature 
shows the extreme compatibility of antiferromagnetic fluctuations and 
$d_{x^2 - y^2}$ superconductivity. To a first approximation, 
the exponent of the power-law decay of the spin-spin correlations 
varies with the coupling strength $W/t$. At $W/t = 0.6$ the QMC data
for  lattice sizes ranging from $L = 4 $ to $ L = 16$  is consistent with
real space spin-spin correlations of the form
$ \cos(\vec{Q} \vec{r}) / | \vec{r} | ^{-\alpha} $ 
with $\alpha  \sim 1.16 $. The QMC data suggests that $\alpha$
decreases continuously to $\alpha = 1$ as the coupling 
$W/t $ approaches $W_c/t$.   Hence,  $W_c/t \sim 0.3$ terminates 
a critical line in the schematic phase diagram drawn in  
Fig. \ref{croc.fig}a.

As a natural consequence of Fig. \ref{croc.fig}a  we expect at 
$ W < W_c$ a doping induced quantum transition between an
antiferromagnetic Mott insulator and $d_{x^2 - y^2}$ superconductor.
As mentioned in the introduction,
the anomalous compressibility and unusually large
dynamical exponent observed numerically at the filling controlled
metal-insulator transition \cite{Imada,Furukawa,Assaad96}. 
enhances the sensitivity of the system to two-particle processes
as generically contained in the form $H_W$. 
Work along those lines, for the filling controlled transition,
is under progress.

To determine the order of the phase transition at $W_c/t$ is presently
beyond the reach of our numerical calculations. 
At the mean field level, 
the transition is expected to be of first order since the phases have 
different broken symmetries. In this scenario, the staggered moment 
would be expected to show a jump at the critical coupling constant. 
Within our numerical accuracy, we do not observe such a  
feature, and the possibility of a continuous phase transition remains open.
An example of a continuous phase transition between two broken 
symmetry states on a square lattice is found in the case of quantum 
antiferromagnets where the Berry phase is a dangerously 
irrelevant operator which leads to 
spin-Peierls ordering (i.e. broken lattice symmetry) in the disordered state
for half-integer spin \cite{Sachdev,Read,Note1}.  

In the $\kappa-$type $BEDT-TTF$ compounds, the antiferromagnetic and
superconducting phases are adjacent to each other in the plane of 
temperature and either pressure, anion substitution or deuteration of
hydrogen atoms \cite{Miyagawa,Kawamoto,Ito}. 
Thus both changes in the bandwidth and in the interaction strength have been
studied.  The insulating phase  of the $\kappa-$type 
$BEDT-TTF$ compounds is a correlated insulator in the sense that
band-structure calculations predict a metal \cite{Williams}. 
A dimer model has been proposed 
to account for the magnetic insulating phase \cite{Miyagawa}. 
Here, a pair of $BEDT-TTF$  
molecules carries a charge of unity, and constitute a single site in terms 
of a Hubbard model.  The {\it on-site}  Hubbard interaction $U_{\rm dimer}$
depends upon the intra-dimer 
hopping, $t_{\rm dimer}$.  In the limit of large Coulomb repulsion per 
$BEDT-TTF$ molecule, $ U_{\rm dimer}  \sim  t_{\rm dimer}$. 
Because of the layered structure of this compound, our two-dimensional 
model may offer a simplified description of the system.
In those compounds, the direct transition line between the antiferromagnetic 
and superconducting phases appears to extend to  finite temperatures, thus 
implying a first order phase transition.  A detailed comparison between  theory
and experiment is beyond the scope of this work.  However, 
an interesting point, is that in the superconducting phase both our 
results and  experimental results show a common feature: the peak temperature
in $ {\rm Re} \chi(\vec{q} = 0, \omega = 0 ) $ is higher than the crossover 
temperature in $1/T_1$ (i.e. $T_1^{cr}$ in our notation).
It would be interesting to see whether the 
antiferromagnetic fluctuations are robust (i.e. divergence of 
$ {\rm Re} \chi(\vec{Q}, \omega = 0 )$) in the superconducting phase.
We also note that the  $^{13}C$ nuclear spin relaxation rate in  the 
superconducting phase of the $\kappa-$type $BEDT-TTF$ compounds has been
reported to follow a $T^3$ law \cite{Kanoda} which would be consistent with
a $d_{x^2 - y^2}$ order parameter.

We have studied spin and charge  degrees of freedom at finite 
temperatures in the superconducting phase, at coupling 
strength $W/t = 0.35 $. As schematically drawn in Fig. \ref{croc.fig}b 
we expect 
the occurrence of a finite temperature Kosterlitz-Thouless transition.  
We give numerical evidence that such a transition indeed 
occurs. After extrapolation to the thermodynamic limit we estimate 
$T_{KT} \sim 0.2 t $ at $W/t = 0.35$. 
In the vicinity of the Kosterlitz-Thouless transition
temperature a pseudo-gap in the  one-electron
density of states, $N(\omega)$, appears. On our finite  sized lattices, the 
pseudo-gap feature  appears above $T_{KT}$.
For the three considered lattice, $L = 6$,  $L=8$  and  $L = 10$, 
the estimated superconducting gap  scales as $T_{KT}$ and satisfies 
$\Delta_{sc}/T_{KT} \sim  2.5$.

In the superconducting phase, at $W/t = 0.35$, 
${\rm Re} \chi (\vec{Q}, \omega = 0 )$  diverges in
the low temperature and thermodynamic limits. 
This divergence as a function of temperature is slower than in the
antiferromagnetic Mott insulating state at $W/t = 0$, and is expected to
follow a power-law. 
We may estimate a magnetic scale by the temperature at which
${\rm Re} \chi (\vec{Q}, \omega = 0 ) $  starts to increase. 
At $W/t = 0.35$, we obtain $J \sim 0.5 t$. 
The QMC data is consistent with the vanishing of 
${\rm Re} \chi (\vec{q} = 0, \omega = 0 ) $   in the zero temperature and
thermodynamic limits. $ {\rm Re} \chi (\vec{q}=0, \omega = 0 ) $ shows 
a maximum at $T \sim J$.  A similar although stronger feature is seen in
the Hubbard model (i.e. $W/t = 0$).
From the relaxation rate  $1/T_{1}$, we estimate a cross-over temperature,
$T^{cr}_1$ below which $1/T_1$ decreases. This temperature scale 
satisfies $T^{cr}_1/T_{KT}  \sim 0.25 $.  In the temperature range
$ T^{cr}_1 < T < J $, $1/T_1$ grows with decreasing temperature. 
At  the temperature scale set by  $T^{cr}_1$ spectral weight
in $S(\vec{Q}, \omega )$ is shifted from low frequencies to form
a peak at $\omega_0$. The location of the peak, $\omega_0$,
scales with $T_{KT}$ as a function of system size and satisfies 
$\omega_0/T_{KT} \sim 0.4 $. 
The spectral weight under this peak diverges in the zero temperature 
and thermodynamic limits.

The superconducting state at $W/t = 0.35$ is thus best characterized by
i) the 
divergence of  $ {\rm Re} \chi(\vec{Q}, \omega=0) $, but the vanishing
of $1/T_1$ in the low temperature and thermodynamic limits 
and ii) $\omega_0/\Delta_{sc} \sim 0.15 << 1  $. 
This behavior is non-trivial and points out to the extreme compatibility
of $d_{x^2-y^2}$ superconductivity and 
antiferromagnetic spin fluctuations. 

To conclude, the here presented model, shows a rich phase
diagram. The doping of the 
model offers the possibility of reproducing some of the features of the 
cuprates at an energy scale accessible to numerical simulations.  \\
$ ^{\star} $ Future address:  Institut f\"ur Theoretische Physik III,
 Universit\"at Stuttgart, Pfaffenwaldring 57, D-70550 Stuttgart, Germany.

\section*{Acknowledgements}
We thank N. Bulut, T. Dahm, W. Hanke,  M. Randeria, S. Sachdev, 
A. Sandvik  and S.C. Zhang  for instructive conversations. 
The numerical simulations were  carried out on the FACOM 
VPP500 at the Supercomputer Center of  the Institute for Solid State
Physics, University of Tokyo. 
F.F.A  thanks the Swiss National Science  foundation for financial support
under the grant number 8220-042824.
D.J.S. and F.F.A. acknowledge partial support from
the National Science Foundation under the grant No. DMR95-27304.
M.I. thanks the Institute of Theoretical
Physics at the University of California, Santa-Barbara for hospitality.
M.I. acknowledges financial support from the National Science Foundation 
under the grant number PHY94-07194 as well as from a Grant-in-Aid for 
Scientific Research on the  Priority Area "Anomalous Metallic
States near the Mott Transition"  and "Novel Electronic  States in Molecular 
Conductors" from the Ministry of Education, Science and Culture, Japan. 

\appendix
\section{Trial Wave Functions for the PQMC}
In this appendix we show how to construct trial wave functions which
generate no sign problem  if $N^{\uparrow} + N^{\downarrow} = N$, and
in the special case, $N^{\uparrow} = N^{\downarrow} = N/2$
are  spin singlets. We provide explicit forms of the trial wave
functions used in this work.
Our starting point is the  non-interaction Hamiltonian:
\begin{equation}
  H_0 = \sum_{\vec{i},\vec{j}} c^{\dagger}_{\vec{i}} 
          T_{\vec{i},\vec{j}}   c_{\vec{j}}.
\end{equation}
$H_0$ corresponds to the Hamiltonian  (\ref{tUW}) in the limit $U = W = 0$.
We have omitted the spin index, and consider the boundary condition given
by (\ref{Bound}).
Let $U$ be the unitary transformation which diagonalizes $T$:
$U^{\dagger} T U = {\rm diag} \left( \lambda_1 \dots \lambda_N \right).$
We define the hole operators by $ h_{\vec{i}} = (-1)^{\vec{i}}
c^{\dagger}_{\vec{i}} $  as well as 
\begin{equation}
\gamma_{\vec{k}}=  \sum_{\vec{i}} U^{\dagger}_{\vec{k},\vec{i}} c_{\vec{i}}
\; \;  {\rm and} \;  \;
\eta_{\vec{k}}=  \sum_{\vec{i}} h_{\vec{i}} (-1)^{\vec{i}} U_{\vec{i},\vec{k}}.
\end{equation}
We have used the notation $ (-1)^{\vec{i}}= (-1)^{i_x + i_y}$. The above 
operators satisfy 
\begin{equation}
[H_0,\gamma^{\dagger}_{\vec{k}}] = \lambda_{\vec{k}} \gamma^{\dagger}_{\vec{k}}
\; \;  {\rm and} \;  \;
[H_0,\eta^{\dagger}_{\vec{k}}] = -\lambda_{\vec{k}} \eta^{\dagger}_{\vec{k}}.
\end{equation}
We set the trial wave function in the up spin sector  to be the
ground state of  $H_0$:  $ | \Psi_T^{\uparrow} \rangle  =
\prod_{\vec{k} \epsilon A  }   \gamma^{\dagger}_{\vec{k}}  
| 0 \rangle $,
where $ | 0 \rangle $ is the electron vacuum  and $ A $ is set of 
$\vec{k}$ points defined by the requirement that $ | \Psi_T \rangle $ is
the ground state of  $H_0$. By construction $ | \Psi_T^{\uparrow} \rangle  
= \prod_{\vec{k} \epsilon \overline{A}}   \eta^{\dagger}_{\vec{k}}  
| 0^h \rangle $, where $| 0^h \rangle $ is the hole vacuum  and 
$ \overline{A} $ is the complement of $A$.
We may hence define:
\begin{eqnarray}
\label{Trial2}
    & & P^{\uparrow}_{ \vec{i}, \vec{k} }  = U_{ \vec{i}, \vec{k} } \; \; \; \;
  \vec{i}: 1 \dots N,  \; \; \vec{k} \epsilon A \nonumber \\
    & & P^{\downarrow}_{\vec{i}, \vec{k} }  = (-1)^{\vec{i}}
         \overline{U}_{\vec{i}, \vec{k}}   \; \; \; \;  
  \vec{i}: 1 \dots N,  \; \; \vec{k} \epsilon \overline{A}.
\end{eqnarray} 
With this choice of $P^{\sigma}$ no sign problem occurs 
(see Eqs.(\ref{Trial}), (\ref{Sign1}), (\ref{Sign2}) and (\ref{Sign3})). 
 We note that 
the only restriction on the particle number 
is $N^{\uparrow} + N^{\downarrow} = N$. 

In the special case $ N^{\uparrow}  = N^{\downarrow} = N/2 $   the trial wave
function $ | \Psi_T \rangle   = 
| \Psi_T^{\uparrow} \rangle  \otimes | \Psi_T^{\downarrow} \rangle$ satisfies:
\begin{eqnarray}
    \sum_{\vec{i}, \vec{j}} 
    \langle \Psi_T | \vec{S}_{\vec{i}} \cdot  \vec{S}_{\vec{j}} | \Psi_T \rangle
  & = &
 \frac{1}{2} \sum_{\vec{k}} f^{\gamma} (\vec{k}) f^{\eta} (\vec{k}) +
          \left( f^{\gamma} (\vec{k}) - 1 \right) 
          \left( f^{\eta} (\vec{k}) - 1 \right)   \; \; \; {\rm where} 
     \nonumber  \\
f^{\eta} (\vec{k}) &  = & \langle \Psi_T^{\downarrow} | \eta^{\dagger}_{\vec{k}} 
 \eta_{\vec{k}}  |  \Psi_T^{\downarrow} \rangle  \; \; {\rm and}  \; \; 
  f^{\gamma} (\vec{k}) = \langle \Psi_T^{\uparrow} | \gamma^{\dagger}_{\vec{k}}
 \gamma_{\vec{k}}  |  \Psi_T^{\uparrow} \rangle.
\end{eqnarray}
If the ground state of $H_0$ with $N/2$ particles per spin sector is 
non-degenerate, then $ \sum_{\vec{i}, \vec{j}}
\langle \Psi_T | \vec{S}_{\vec{i}} \cdot  \vec{S}_{\vec{j}} | \Psi_T \rangle $ 
vanishes, independently of the choice of the unitary transformation $U$. 
For non-zero values of the flux $\Phi$ 
(i.e. $\Phi \neq 2 \pi n \Phi_0$, $n$ being an integer), 
this condition is satisfied, and hence 
a trial wave function showing no sign problem and satisfying (\ref{Trial1}),
may easily be constructed numerically.

\begin{figure}
\epsfbox{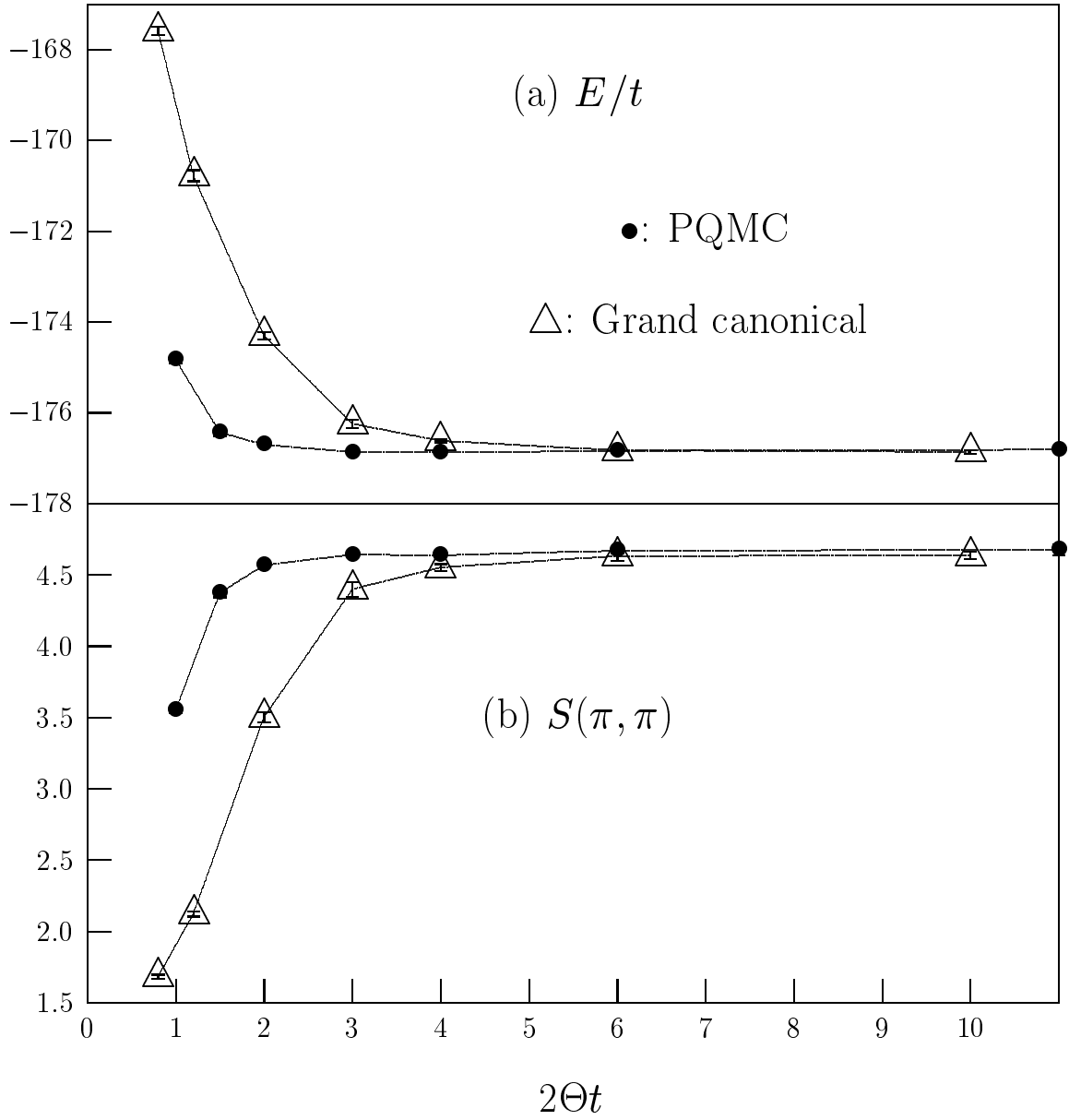}
\mbox{}\\[0.5cm]
\caption[]
{(a) Total energy $E$  and  (b) spin structure factor at $\vec{Q} = (\pi, \pi)$
as obtained from the finite temperature QMC algorithm at $\beta = 2 \theta$
and from the PQMC algorithm. 
\label{theta.fig} }
\end{figure}

At $\Phi = 0$,  the  ground state of $H_0$ with $N/2$ particles  per spin sector
is degenerate, and the total spin of the trial wave function 
ill depend on the
choice of the unitary transformation $U$.  In principle, one can use
the trial wave function obtained at an {\it infinitesimal} value of 
$\Phi$ to lift the degenerancy.  This procedure  would  force us to work with 
complex numbers which enhances the CPU time  by a factor three to four. To 
easily circumvent those problems we define a new Hamiltonian 
\begin{eqnarray}
\label{Dimer}
 \tilde{H}_0 =& &  -t ( 1 - \delta)  \sum_{\vec{i}} 
 \left( c^{\dagger}_{\vec{i}}  c_{\vec{i} + \vec{a}_y}  + {\rm h.c.} \right)
-t ( 1 - \delta)  \sum_{\vec{i} = (2 i_x +1, i_y) }
  \left( c^{\dagger}_{\vec{i}}  c_{\vec{i} + \vec{a}_x}  + {\rm h.c.} \right)
\nonumber \\ & & -t ( 1 + \delta)  \sum_{\vec{i} = (2 i_x , i_y) }
  \left( c^{\dagger}_{\vec{i}}  c_{\vec{i} + \vec{a}_x}  + {\rm h.c.} \right)
\end{eqnarray}
with periodic boundary conditions in both lattice directions. 
For {\it infinitesimal} values of $\delta$, the ground state of $\tilde{H}_0$ 
with $N/2$ particles is non-degenerate.  Diagonalizing $\tilde{H}_0$ generates
an orthogonal matrix $U$ from which one may construct the trial wave 
function.  
Both for the here presented model (see Fig. \ref{theta.fig}) and for the 
half-filled Hubbard model \cite{Assaad96a}, 
the above trial wave functions provide quick convergence to the ground state.
The reason lies in the fact that they are spin singlets and hence orthogonal
to spin-wave excitations which constitute the low energy part of the 
spectrum at least in the case of the half-filled Hubbard model.

\unitlength5mm

\newpage

\newpage

\newpage

\newpage

\newpage

\newpage

\newpage

\newpage

\newpage

\newpage

\newpage

\newpage

\newpage

\begin{thebibliography}{99}
\bibitem{Scalapino94} For a review see 
D.J. Scalapino {\it Does the Hubbard model have the right
stuff?} Proceedings of the International School of Physics, Enrico Fermi,
Course CXXI,   edited by R.A. Broglia and J.R. Schrieffer, North-
Holland, (1994),  and references therein.

\bibitem{Assaad97} F.F. Assaad, M. Imada and D.J. Scalapino, Phys. Rev. Lett.
{\bf 77}, 4592,  (1996).

\bibitem{Su} W.P. Su, J.R. Schrieffer, and A.J. Heeger, Phys. Rev. B
{\bf 22}, 2099 (1980).

\bibitem{Hirsch1} J.E. Hirsch, Phys. Rev. B{\bf 35}, 8726, (1987).

\bibitem{Imada1} M. Imada, Prog. Theor. Phys. Suppl. {\bf 113}, 203, (1993).

\bibitem{Imada} M. Imada,  J. Phys. Soc. of Jpn. {\bf 64}, 2954 (1995).

\bibitem{Furukawa}  N. Furukawa and M. Imada, J. Phys. Soc. Jpn. {\bf 62},
2557, (1993). N. Furukawa, F.F. Assaad and  M. Imada, J. Phys. Soc. Jpn
{\bf 65}, 2339, (1996).

\bibitem{Assaad96}  F.F. Assaad and M. Imada, Phys. Rev. Lett. {\bf 76},
3176, (1996).

\bibitem{Bulut93} N. Bulut, D.J. Scalapino and S.R. White 
Phys. Rev. B {\bf 47}, 6157, (1993), Phys. Rev. B {\bf 47}, 14599, (1993).

\bibitem{Blankenbecler}  R. Blankenbecler and R.L. Sugar,  Phys. Rev.  D 
{\bf 27}, 1304 (1983).

\bibitem{Koonin} G. Sugiyama and S.E. Koonin, Annals of Phys.{\bf 168}
(1986) 1.

\bibitem{Sandro} S. Sorella, S. Baroni, R. Car, And M. Parrinello,
Europhys. Lett. {\bf 8} (1989) 663.
S. Sorella, E. Tosatti, S. Baroni, R.
Car, and M. Parinello, Int. J. Mod. Phys. B{\bf 1} (1989) 993.

\bibitem{Imada89} M. Imada and Y. Hatsugai, J. Phys. Soc. Jpn. {\bf 58}, 3752,
(1989).

\bibitem{Hirsch85} J.E.Hirsch, Phys. Rev. B {\bf 31}, 4403, (1985).
\bibitem{White} S.R. White et al.  Phys. Rev. B{\bf 40}, 506, (1989).


\bibitem{Assaad94b} F.F. Assaad, W. Hanke and D.J. Scalapino, 
Phys. Rev. B  {\bf 50}, 12835, (1994)

\bibitem{Assaad96a} 
 F.F. Assaad and M. Imada, J. Phys. Soc.  Jpn. {\bf 65},189, (1996). 


\bibitem{Fye}  R.M. Fye, Phys. Rev B{\bf 33}, 6271 (1986).

\bibitem{Hirsch83} J.E.Hirsch, Phys. Rev. B {\bf 28} (1983) 4059.

\bibitem{Loh} E. Loh and J. Gubernatis, in Modern Problems of
Condensed  Matter Physics, edited by W. Hanke and Y.V. Kopaev,
(North Holland, Amsterdam, 1992), Vol 32, p. 177.


\bibitem{Jarrel} M. Jarrel and J.E. Gubernatis, 
Physics Reports, {\bf 269}, (1996) 133

\bibitem{Linden} W. von der Linden, 
Applied Physics A{\bf60},   (1995), 155.


\bibitem{Kohn} W. Kohn, Phys. Rev. {\bf 133}, A171, (1964).

\bibitem{Byers} N. Byers and C.N. Yang, Phys Rev. Lett.{\bf 7}, 46, (1961).

\bibitem{Yang}  C.N. Yang, Reviews of Mod. Phys. {\bf 34}, 694 (1962).

\bibitem{Assaad93} F.F. Assaad, W. Hanke and D.J. Scalapino, 
Phys. Rev. Lett. {\bf 71}, 1915 (1993).  

\bibitem{KT} J.M. Kosterlitz and D.J. Thouless, J. Phys. C. {\bf 6},
1181, (1973).

\bibitem{KT1} D.R. Nelson and J.M. Kosterlitz, Phys. Rev. Lett. {\bf
39}, 1201 (1977).

\bibitem{Himbergen} J.E. Van Himbergen and S. Chakravarty, Phys. Rev. B
{\bf 23} 359 (1981).

\bibitem{Loh85} E. Loh, Jr., D.J. Scalapino and P.M. Grant, Phys. Rev. B
{\bf 31}, 4712 (1985).

\bibitem{Assaad94c} F.F. Assaad, W. Hanke and D.J. Scalapino,  Phys. Rev. B
{\bf 49}, 4327 (1994).

\bibitem{White89} S.R. White, D.J. Scalapino, R.L. Sugar, N.E.
Bickers and R.T. Scalettar,  Phys. Rev. B{\bf 39}, 839, (1989).

\bibitem{Nejat} N. Bulut and D.J. Scalapino, Phys. Rev. B {\bf 45} 2371,
(1992).

\bibitem{Chakravarty} S. Chakravarty , B.I. Halperin and D.R. Nelson, Phys.
Rev. Lett. {\bf 60}, 1057 (1988); Phys. Rev. B{\bf 39}, 2344 (1989).

\bibitem{Sachdev} A. Chubukov, S. Sachdev and J. Ye, Phys. Rev. B {\bf 49},
11919 (1994).  

\bibitem{Read} N. Read and S. Sachdev, Phys. Rev. B {\bf 42} 4568, (1990).

\bibitem{Note1} Although we have used the SU(2) language, this result is
derived in the large N limit of SU(N) quantum antiferromagnets. 

\bibitem{Miyagawa} K. Miyagawa,A. Kawamoto,  Y. Nakazawa and K. Kanoda,
Phys. Rev. Lett {\bf 75}, 1174, (1995). 
\bibitem{Kawamoto} A. Kawamoto, K. Miyagawa, Y. Nakazawa and K. Kanoda,
Phys. Rev. B {\bf 52} 15522 (1995).
\bibitem{Ito} H. Ito, T. Ishiguro, M. Kubota and G. Saito, 
J. Phys. Soc. Jpn. {\bf 65} 1987 (1996).
\bibitem{Williams} J.M. Williams et. al. Science {\bf 252} 1501 (1991),
and references therein. 
\bibitem{Kanoda} K. Kanoda, K. Miyagawa, A. Kawamoto and Y. Nakazawa,
Phys. Rev B{\bf 54}, 76, (1996).

\end{thebibliography}
\end{document}